\documentclass[11pt,preprint]{aastex}

\usepackage{color}
\usepackage{natbib}
\usepackage{amsmath}
\usepackage{threeparttable}
\usepackage{multirow}
\usepackage{listings}
\usepackage{url}
\usepackage[normalem]{ulem} 

\tightenlines
\slugcomment{}

\newcommand \beq        {\begin{equation}}
\newcommand \beqa	{\begin{eqnarray}}

\newcommand \eeq	{\end{equation}}
\newcommand \eeqa	{\end{eqnarray}}

\newcommand \gtsim	{\gtrsim}		 

\newcommand{\oldtext}[1]{}

\countdef\decade=200
\advance\decade by \year
\countdef\hours=201
\advance\hours by \time
\divide\hours by 60
\countdef\mins=202
\advance\mins by \hours
\multiply\mins by 60
\multiply\hours by 100
\countdef\miltime=203
\advance\miltime by \hours
\advance\miltime by \time
\advance\miltime by -\mins

\renewcommand{\Im}[1]{{\rm Im}\left(#1\right)}
\renewcommand{\Re}[1]{{\rm Re}\left(#1\right)}
\renewcommand{\vec}[1]{\mathbf{#1}}

\renewcommand{\O}[1]{\mathcal{O}\left(#1\right)}
\newcommand \nhat {\mathbf{\hat{n}}}

\newcommand \aee {a_{\rm EE}}
\newcommand \aeff {a_{\rm eff}}
\newcommand \Aproj {A_{\rm proj}}
\newcommand \Vtruei {V_i^{\rm true}}   
\newcommand \sigextfit {\sigma_{\rm ext}^{\rm fit}}
\newcommand \newedit[1]{#1}
\newcommand \refereenote[1]{}
\begin{document}

\title{ Accurate Modeling of X-ray Extinction by Interstellar Grains}

\author{
John Hoffman\altaffilmark{1}
and
B.\ T.\ Draine\altaffilmark{1}
}
\altaffiltext{1}{%
Princeton University Observatory, Peyton Hall, Princeton, NJ 08544-1001, USA;
jah5@astro.princeton.edu, draine@astro.princeton.edu}

\begin{abstract}

Interstellar abundance determinations from fits to X-ray absorption
edges often rely on the incorrect assumption that scattering is insignificant 
and can be ignored. We show instead that scattering contributes 
significantly to the attenuation of X-rays for realistic dust grain 
size distributions and substantially modifies the spectrum near 
absorption edges of elements present in grains. The dust
attenuation modules used in major X-ray spectral fitting programs do
not take this into account. We show that the consequences of
neglecting scattering on the determination of interstellar elemental
abundances are modest; however, scattering (along with uncertainties
in the grain size distribution) must be taken into account when
near-edge extinction fine structure is used to infer dust
mineralogy. We advertise the benefits and accuracy of anomalous
diffraction theory for both X-ray halo analysis and near edge
absorption studies. An open source Fortran suite, General Geometry
Anomalous Diffraction Theory (GGADT), is presented that calculates
X-ray absorption, scattering, and differential scattering cross
sections for grains of arbitrary geometry and composition.

\end{abstract}
\keywords{dust, extinction, scattering, X-rays: general, X-rays: ISM, ISM: abundances}

\section{Introduction
         \label{sec:intro}}
The absorption and scattering of light by interstellar dust and gas
has been of interest to astronomers for over a century. Evidence for
dust existing between stars in the Milky Way first appeared in the
astronomical literature when \cite{Herschel_1785} described gaps in
the density of stars across the sky. The notion that interstellar dust
was responsible for the dimming of starlight appears to have been
first proposed by \cite{Struve_1847} and independently by
\cite{Pickering_1897},
\cite{Kapteyn_1904,Kapteyn_1909a,Kapteyn_1909b}, and
\cite{Barnard_1907,Barnard_1910}.

Since the early 20th century, our understanding of the dust in the
interstellar medium (ISM) has continued to evolve. Extensive studies
of the wavelength-dependence of extinction (i.e., reddening)
established that submicron grains are present in the
ISM. \cite{Schalen_1938} estimated a characteristic radius of
$\sim0.05\mu$m, and detailed calculations by
\cite{Oort+vandeHulst_1946} later put the characteristic radius of ISM
dust grains at $\sim0.3\mu$m. In 1949, \cite{Hall_1949} and
\cite{Hiltner_1949a,Hiltner_1949b} discovered the polarization of
starlight. Their discovery was the first evidence that ISM dust grains
are (1) non-spherical, and (2) coherently aligned over large distance
scales. In recent decades, the study of interstellar grains has used
observations of emission, absorption, and scattering, ranging from
microwaves to X-rays.

Interstellar grains absorb and scatter X-rays. Both gas and dust
attenuate X-rays propagating through the ISM, but elemental X-ray
absorption edges differ between atoms, ions, and solids, so that
near-edge X-ray absorption fine structure (NEXAFS) can in principle
reveal the composition of interstellar grains
\citep{Martin_1970,Martin+Sciama_1970,Evans_1986,Woo_1995,Forrey_et_al_1998,Draine_2003c,Lee+Ravel_2005,Lee_et_al_2009,Constantini_et_al_2012,Pinto_et_al_2013} 

X-ray halos surrounding astrophysical sources provide another valuable
tool with which to study the ISM
\citep{Hayakawa_1970,Martin_1970}. Small angle scattering of X-rays by
dust grains along the line of sight produces a halo around the source
\citep{Overbeck_1965,Hayakawa_1970,Martin_1970}. Measurements of these
X-ray halos can be used to test and constrain dust models \citep[see
  e.g.][]{Smith+Dame+Costantini+Predehl_2006}. Recent work by \cite{Seward+Smith_2013}
  used \emph{Chandra} observations of Cyg X-1 to
  look for azimuthal asymmetry in the surrounding X-ray halo, a technique
  that could potentially be used to constrain dust shape and grain alignment.
 Observations of X-ray
halos around variable sources can constrain the orientation and
geometry of dust clouds, allowing astronomers to study the ISM in
three dimensions
\citep{Predehl+Burwitz+Paerels+Truemper_2000,Vaughan+Willingale+Obrien+etal_2004,Tiengo_et_al_2010,
  Heinz_et_al_2015}, and could even be used for extragalactic distance
determination \citep{Draine+Bond_2004}.

The focus of this paper will be on the importance of accounting for
X-ray scattering when inferring abundances of different grain
materials from absorption edge measurements. The
\cite{Wilms+Allen+McCray_2000} model for X-ray attenuation, which is
employed by XSPEC \newedit{\citep{Arnaud_1996}}, Spex \newedit{\citep{Kaastra+Mewe+Nieuwenhuijzen_1996}}, 
and several other X-ray data analysis codes,
does not include scattering; this, together with an approximate treatment
of absorption in large grains that assumes each grain is a slab of thickness $4a/3$, 
can result in incorrect conclusions regarding the composition and abundance 
of interstellar dust.

The Rayleigh-Gans (RG) approximation \citep{Mauche+Gorenstein_1986},
often used to model X-ray halos, is also prone to errant
application. As shown in \cite{Smith+Dwek_1998}, for an MRN
\citep{Mathis+Rumpl+Nordsieck_1977} size distribution of spherical
graphite and silicate grains, the RG approximation substantially
overestimates the intensity of the soft X-ray ($\lesssim 1$ keV)
scattering halo. Anomalous diffraction theory (ADT), by contrast, is
accurate and easy to use.

The paper is organized as follows: in section
\ref{sec:modeling:overview}, the problem of modeling dust extinction
is discussed, along with popular theoretical and computational
techniques for solving the scattering problem. Section
\ref{sec:modeling:misuse} explains the perils of ignoring dust
scattering when modeling observations of X-ray extinction, especially
when inferring elemental abundances and mineralogy from X-ray
absorption edges. In section \ref{sec:ggadt}, the advantages of using
ADT to model dust scattering and absorption are discussed. An open
source code suite, ``GGADT,'' which uses ADT to calculate scattering
and absorption by grains of arbitrary composition and geometry is
presented in section \ref{sec:ggadt:code}. Section
\ref{sec:discussion} summarizes the salient points discussed in this
paper. \newedit{More details about GGADT are given in Appendices
\ref{ggadt_desc}, \ref{examples_ggadt}, \ref{efficient_calc_ggadt}, and
\ref{efficient_calc_ggadt2}}.

\section{Modeling X-ray extinction by ISM dust
		\label{sec:modeling}}

\subsection{Overview of Theoretical and Numerical Techniques
	\label{sec:modeling:overview}}

\begin{figure}[t]
\centering
\includegraphics[width=0.9\linewidth]{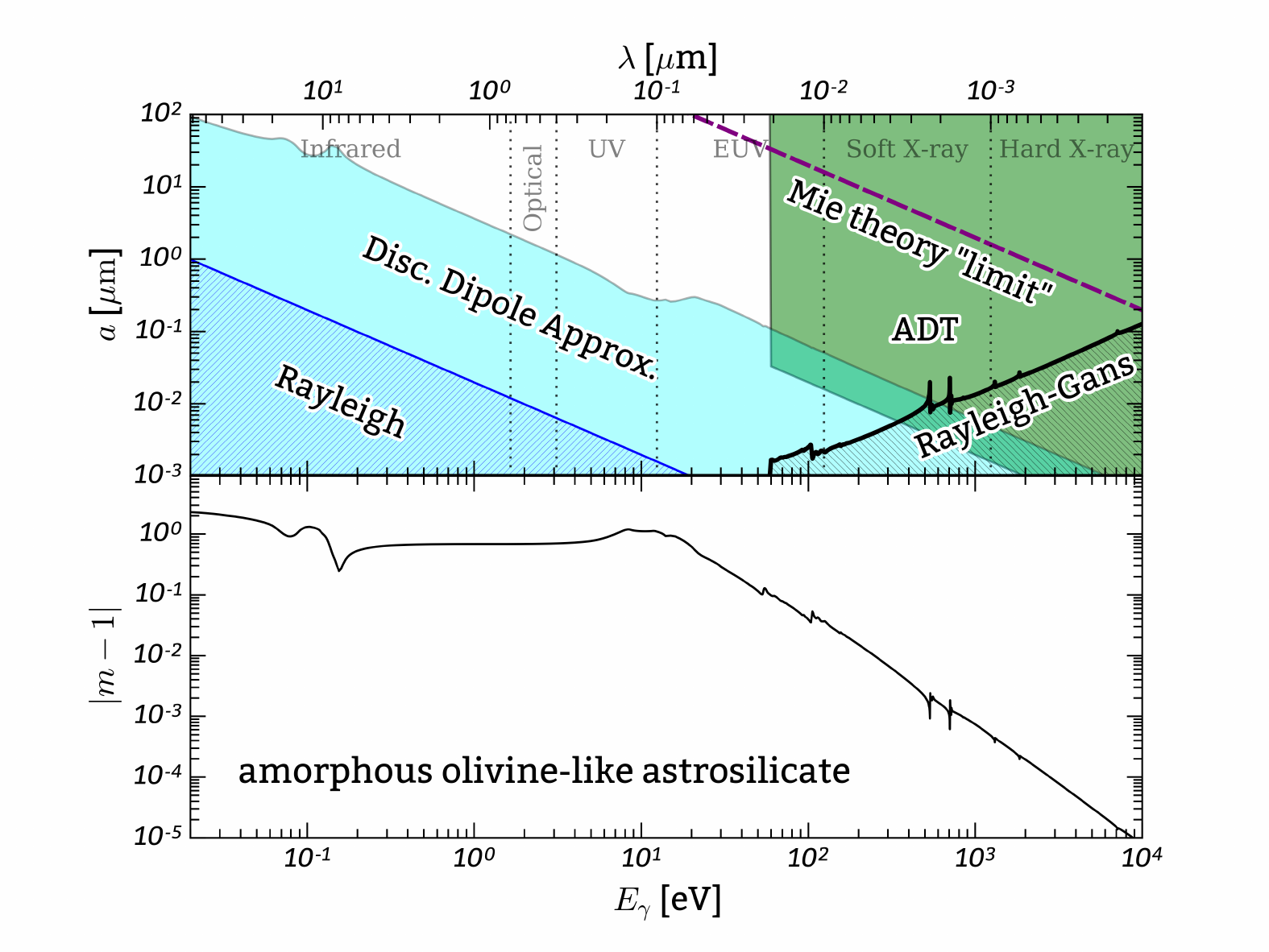}
\caption{\footnotesize
  \label{fig:regions_of_validity} Upper panel: Validity regions for four
  popular approximation schemes for calculating scattering and absorption
  by nonspherical grains (see text). Though Mie theory is only applicable 
  to spherical grains, we plot a line corresponding to $x\equiv ka = 10^4$, above 
  which numerical implementations of Mie theory become prone to 
  round-off errors.
  At X-ray energies, anomalous diffraction theory (ADT) is the
  method of choice except for extremely small grains, where the
  discrete dipole approximation (DDA) or the Rayleigh-Gans approximation
  can be used.
  Note invalidity of the Rayleigh-Gans approximation for grain
  sizes larger than $\sim$0.02$\mu{\rm m}\cdot(E/{\rm keV})$.
  Lower panel: $|m-1|$ for silicate material.}
\end{figure}

For spheres, Mie theory provides a truncated multipole expansion of
the full solution to the scattering problem. As grain radius $a$
increases, the number of multipole terms required also increases. For
grain sizes that are much larger than the incident wavelength
$\lambda$, Mie theory becomes computationally demanding, and computer
implementations of Mie theory are limited by roundoff error. Codes are
available that can handle size parameters $x\equiv 2\pi a/\lambda$ as
large as $\sim10^4$ \citep{Wiscombe_1980}. 
However, $x=5070 (a/\mu{\rm m})(E/{\rm keV})$ can
exceed this limit for large grains at high energies. Mie theory is 
limited to spheres, but the polarization of starlight
indicates that interstellar grains are not spherical 
\citep{Hall_1949,Hiltner_1949a,Heiles_2000} and thus other methods 
are needed to model the extinction of light by interstellar \newedit{dust}.

Several approximation schemes have been used to calculate the scattering
and absorption of light by non-spherical grains. Among the more
popular methods are: the electric dipole approximation (Rayleigh
scattering), the Rayleigh-Gans approximation, anomalous diffraction
theory (ADT), and the discrete dipole approximation
(DDA). \newedit{For a given material composition,} each of these 
\newedit{\sout{schemes} approximations} is valid for a range of grain sizes and
electromagnetic wavelengths. The domains of validity for astrosilicate
grains are shown in Figure \ref{fig:regions_of_validity}. 

The Rayleigh-Gans approximation is often used in X-ray astronomy to
model X-ray halos [see,
e.g., \cite{Vaughan+Willingale+Obrien+etal_2004}]. The Rayleigh-Gans
approximation assumes that each infinitesimal volume element of the
grain responds only to the incident electric field. The total
scattered field produced by the entire grain is computed by
integrating over the dipole scattering contributions from all volume
elements.

The Rayleigh-Gans approximation assumes (1) $|m-1| \ll 1$, where $m$
is the complex index of refraction of the grain material, and (2)
$2ka|m-1|\ll 1$, negligible complex phase shifts in the incident wave
as it travels through the grain, where $k\equiv 2\pi/\lambda$. For
astrosilicate dust and X-ray wavelengths, these assumptions hold only
for small grains ($a \lesssim 0.02\mu{\rm m}\cdot(E/{\rm keV})$).

The discrete dipole approximation (DDA)
\citep{Purcell+Pennypacker_1973,Draine+Flatau_1994} discretizes the
grain into a number of finite volume elements, and can be used to
calculate scattering and absorption by grains with arbitrary
geometries. The finite volume elements must be small enough that they
can be treated as dipoles. The DDA, unlike the Rayleigh-Gans
approximation, does not assume that the volume elements are
non-interacting. The DDA is constrained by computational requirements
to problems with $|m|ka \lesssim 30$, which limits it to $a\lesssim
0.006 \mu{\rm m}\cdot({\rm keV}/E)$, hence the DDA is not useful at
X-ray energies.

Anomalous diffraction theory (ADT) \citep{van_de_Hulst_1957} is
applicable to grains of arbitrary geometry that are large compared to
the incident wavelength. The approximations used in the derivation of
ADT require that $|m-1| \ll 1$, and $ka \gg 1$. 
\cite{Draine+Allaf-Akbari_2006} showed that the validity
conditions for ADT ($|m-1| \lesssim 0.1$ and $ka\gtrsim 10$) are
satisfied for silicate grains when $E\gtsim60$eV and
$a\gtrsim0.035\mu$m$\times\left(\frac{60~\text{eV}}{E}\right)$.
The computational requirements are modest, and it is readily
extended to arbitrary geometries.

\newedit{A number of other approximations exist to efficiently 
compute scattering and absorption in various limiting cases.
For optically ``soft'' ($|m - 1| \ll 1$) particles, \cite{Sharma+Somerford_2006}
provides a comprehensive comparison of many available approximations, 
and their accuracy.}

\subsection{Some Widely-Used Models of X-Ray Attenuation by Dust Grains
	\label{sec:modeling:misuse}}

\begin{figure}[!ht]
\centering
\includegraphics[width=0.7\linewidth]{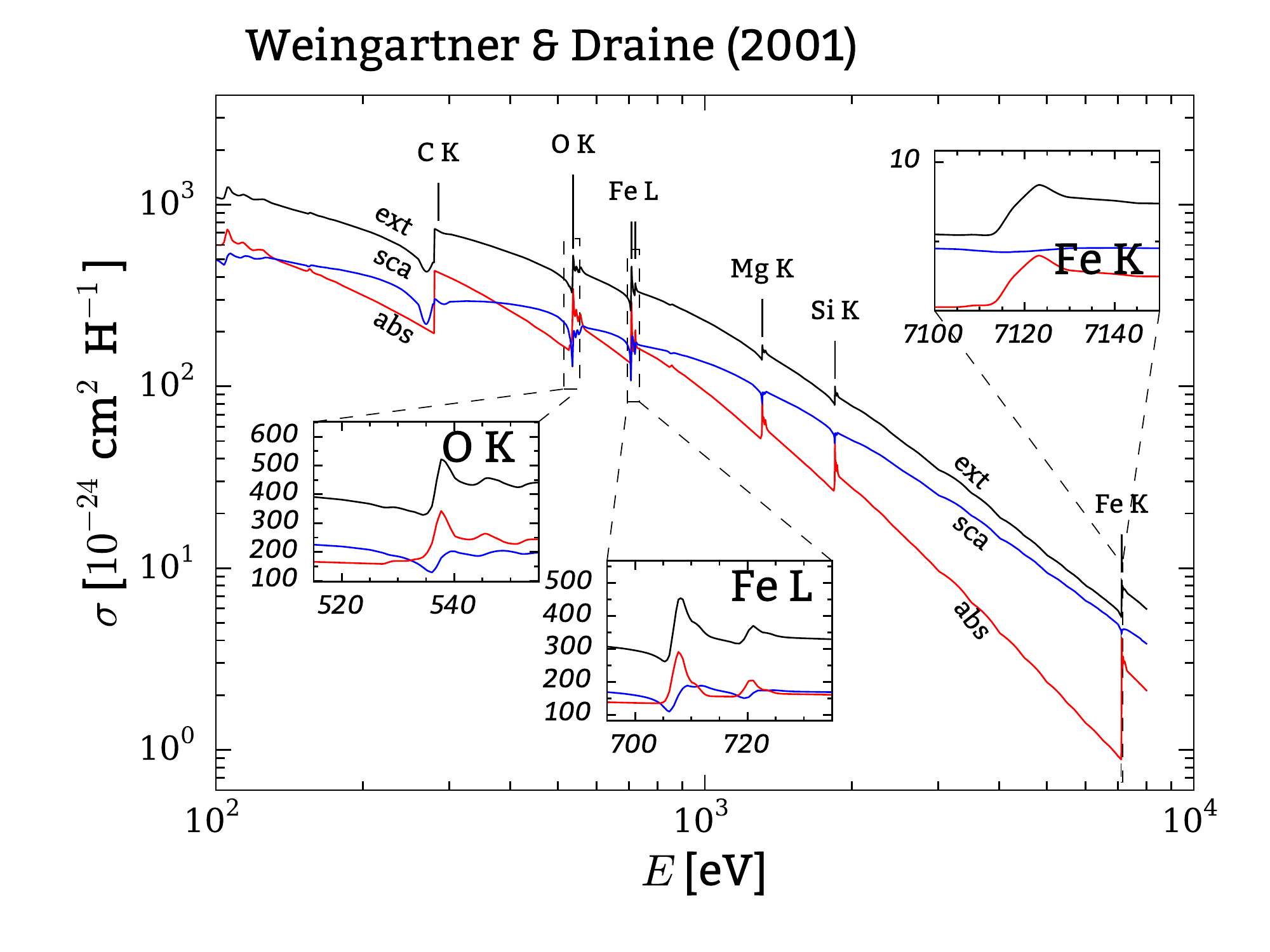}
\caption{\footnotesize\label{fig:extinction_breakdown} Cross sections
  per Hydrogen nucleus for the \cite{Weingartner+Draine_2001a}
  ($R_V=3.1$) dust model. Scattering contributes significantly to the
  extinction, with significant variation across the O K and Fe L
  absorption edges.  }
\end{figure}

\begin{figure}[!ht]
\centering
\includegraphics[width=0.7\linewidth]{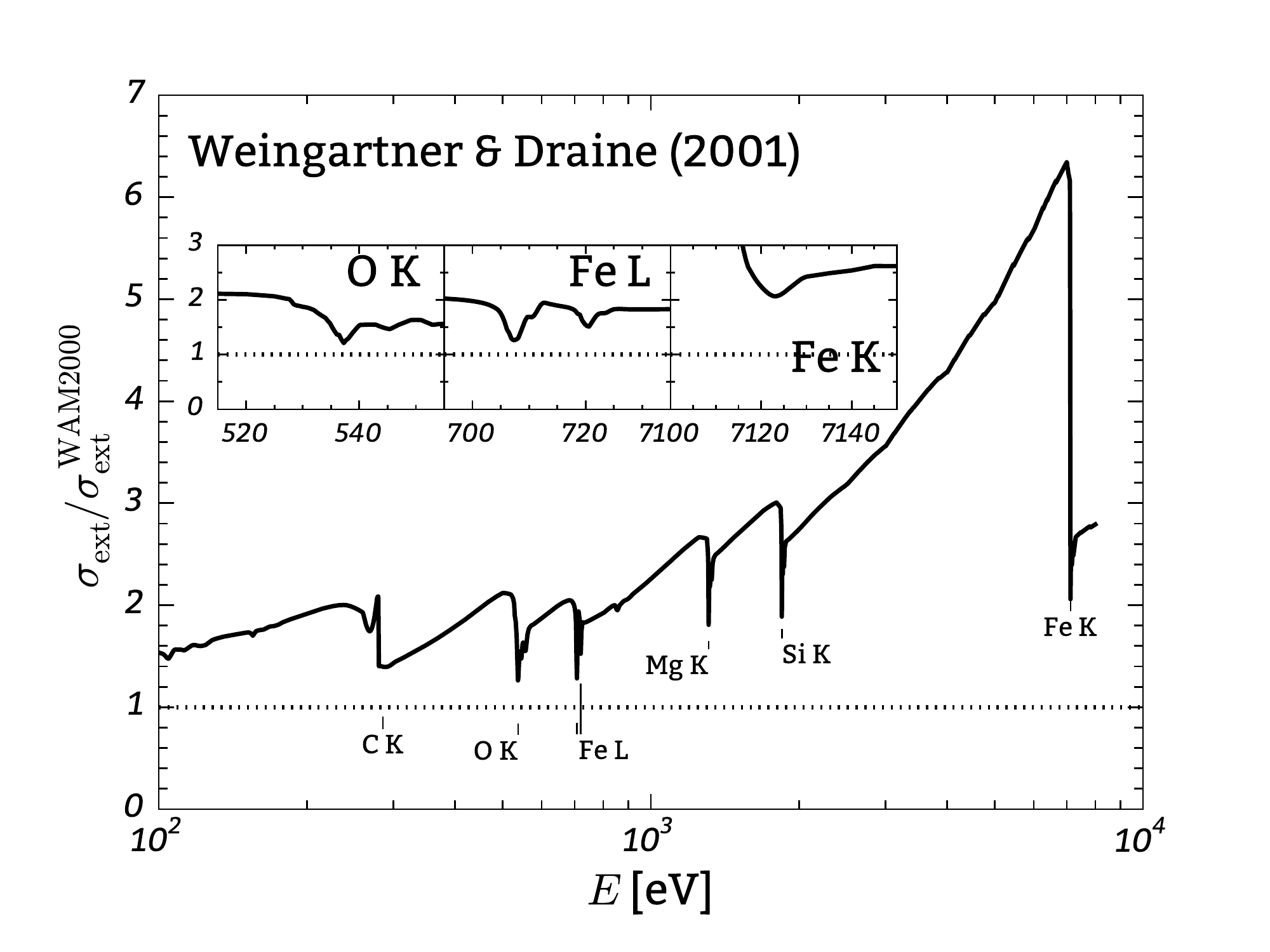}
\caption{\footnotesize\label{fig:alpha_cext_cabs} Exinction cross
  section per Hydrogen nucleus for the WD01 size distribution compared
  to WAM2000 calculations. Though WAM2000 is suitable for estimating
  the \emph{absorption} cross section, scattering contributes
  significantly to extinction, even at high energies. }
\end{figure}

One popular but flawed technique that has been used in the X-ray
astronomy literature is to ignore scattering contributions to the dust
extinction. The \texttt{tbvarabs} routine in the XSPEC
package\footnote{\url{http://heasarc.gsfc.nasa.gov/xanadu/xspec/}} and
the \texttt{dabs} routine in the Spex
package\footnote{\url{http://www.sron.nl/spex}} both use the ISM model
of \cite{Wilms+Allen+McCray_2000} (hereafter WAM2000), which
approximates the extinction cross section of a single grain as the sum
of atomic photoionization cross sections with an approximate correction for
grain self-absorption, and integrates over the dust size distribution
given by \cite{Mathis+Rumpl+Nordsieck_1977} (MRN):
\begin{equation}
\label{wam2000_eq}
\sigma_{\rm ext}^{\rm WAM2000}(E) = \sum_i\int_{a_{\rm
    min}}^{a_{\rm max}} \left(\frac{1}{n_{\rm
    H}}\frac{dn_i}{da}\right)_{\rm MRN} \pi a^2\left(1 -
e^{-\frac{4}{3}\alpha_i(E) a}\right) da,
\end{equation}
where $i$ refers to different grain materials and
$\alpha_i(E)=(4\pi/\lambda)\Im{m_i}$ is the attenuation
coefficient for grain material $i$ with complex refractive index $m_i(E)$.

The WAM2000 attenuation model ignores scattering contributions to the
dust extinction and further approximates a grain of radius $a$ as
having a uniform thickness $(4/3)a$. For sufficiently small grains, a
grain of volume $V$ has an absorption cross section $C_{\rm abs}(E)
\approx \alpha(E)V$ and scattering is unimportant, as assumed by the
WAM2000 dust attenuation model. For larger grains, however, scattering
contributes significantly to the extinction, and should be taken into
account when modeling attenuation.

Figure \ref{fig:extinction_breakdown} shows the contributions from
scattering and absorption to the extinction cross section of a
\citet[hereafter WD01]{Weingartner+Draine_2001a} grain size
distribution. Figure \ref{fig:alpha_cext_cabs} shows the ratio of the
true extinction cross section and the cross section obtained from the
WAM2000 approximation. Figures \ref{fig:extinction_breakdown} and
\ref{fig:alpha_cext_cabs} show that scattering is an important
contributor to grain extinction at X-ray energies for realistic models
of ISM grains.

To illustrate the importance of including scattering when computing
extinction by interstellar dust grains, we use the WAM2000
model to attempt to recover silicate and carbonaceous masses of interstellar dust
from simulated, noise-free observations of extinction.

First, the extinction cross section per H nucleon for ISM dust is calculated for plausible size
distributions (either MRN or WD01):
\begin{equation}
\label{truesigext}
\sigma_{\rm ext}(\lambda) = \sum_i \int_0^{\infty} C_{{\rm
    ext},i}(\lambda, a)\left(\frac{1}{n_{\rm
    H}}\frac{dn_i}{da}\right)_{\!mod}da,
\end{equation}
where $C_{{\rm ext},i}(\lambda,a)$ is the extinction cross section for
a spherical grain composed of material $i$ with radius $a$, and
$(n_{\rm H}^{-1}dn_i/da)_{\!mod}$ is the
size distribution of grain material $i$ for grain model $mod$.

After calculating $\sigma_{\rm ext}(E)$ for a given grain model near an
absorption edge $j$, we imagine that this has been measured (without noise)
and attempt to recover the total volume of
material $i$, $\Vtruei$, by using an attenuation model similar to
WAM2000: near absorption edge $j$ we fit the true $\sigma_{\rm ext}$
from Eq. \ref{truesigext} with
\begin{equation}
\label{eq:sigfit}
\sigextfit = \sum_i \frac{V^{\rm fit}_{ij}}{\Vtruei} \int_{a_{\rm min}}^{a_{\rm
    max}} \left(\frac{1}{n_{\rm H}}\frac{dn_i}{da}\right)_{\!mod}\pi
a^2\left(1 - e^{-\frac{4}{3}\alpha_i(\lambda) a}\right)da + C_j.
\end{equation}
\refereenote{Added ``fit'' superscript to $V_{ij}$.}
The attenuation coefficients $\alpha_i$ are presumed to be known, and
the shape of the size distribution $dn_i/da$ is assumed to be known,
but the multiplier $V^{\rm fit}_{ij}$ and the additive offset $C_j$ are free
parameters: $V^{\rm fit}_{ij}$ is the volume per H of grain material $i$ fit to
absorption edge $j$, and $C_j$ is a constant offset for absorption
edge $j$.\footnote{Because in general we lack a reliable estimate for
  the unattenuated spectrum of the X-ray source, we include an
  adjustable offset $C_j$ for each absorption edge.} The domain of the
fit contains only wavelengths close to absorption edges, and $C_j$ is
fit at each absorption edge.

Fitting is done via the {\tt curve\_fit} function in the {\tt scipy}
Python library \citetext{Jones et al. 2001}\nocite{Scipy}, which
implements the Levenberg-Marquardt non-linear least squares fitting
method \citep{Levenberg_1944,Marquardt_1963}. For each edge, we fit
the \newedit{extinction} over the energy range $E_{\rm edge}\pm \Delta E$. We
tried three values of $\Delta E$ (10, 20, and 30 eV) to investigate
how the results might depend upon the energy range used for the fit.

Two size distributions were considered: the MRN
\citep{Mathis+Rumpl+Nordsieck_1977} dust grain distribution ($dn/da
\propto a^{-3.5}$, $5{\rm nm} < a < 250{\rm nm}$) and the WD01 size
distribution. Both size distributions use spherical grains. We assume
a carbonaceous volume fraction of $f_{\rm carb} = 0.488$ for the MRN
size distribution and $f_{\rm carb} = 0.365$ for the WD01 size
distribution. The refractive indices $m_i$ for both materials are taken from
\citet{Draine_2003c}.

\begin{table}[h]
\footnotesize
\begin{center}
\caption{\label{fit-table} Material volumes $V^{\rm fit}_{ij}$ estimated from WAM2000 fit; $\Vtruei$ is the true volume. }
\begin{tabular}{l|c|l|l|l|l|l}
$j^a$ & Edge & $E_{\rm edge}^b$ & $\Delta E$ & $i^c$    & \multicolumn{2}{c}{$V^{\rm fit}_{ij}/\Vtruei$} \\
\hline \hline
      &      & (eV)              & (eV)        &          &   MRN$^d$   & WD01$^e$  \\
\hline
\hline
\multirow{3}{*}{1}    & \multirow{3}{*}{C K} & \multirow{3}{*}{285.0} 
           & 10.0 &  \multirow{3}{*}{carb.} &
                                          0.952 & 0.798  \\

 &    &  & 20.0 &   &    1.064 & 0.847  \\

 &    &  & 30.0 &   &    0.949 & 0.643  \\

\hline
\multirow{3}{*}{2}    & \multirow{3}{*}{O K} & \multirow{3}{*}{537.0} 
           & 10.0 &  \multirow{3}{*}{sil.} &
                                          1.091 & 1.050  \\

 &    &  & 20.0 &   &    0.935 & 0.911  \\

 &    &  & 30.0 &   &    0.843 & 0.836  \\

\hline
\multirow{3}{*}{3}    & \multirow{3}{*}{Fe L} & \multirow{3}{*}{713.5} 
           & 16.5 &  \multirow{3}{*}{sil.} &
                                          1.084 & 1.011  \\

 &    &  & 26.5 &   &    1.043 & 0.987  \\

 &    &  & 36.5 &   &    0.994 & 0.954  \\

\hline
\multirow{3}{*}{4}    & \multirow{3}{*}{Mg K} & \multirow{3}{*}{1310.0} 
           & 10.0 &  \multirow{3}{*}{sil.} &
                                          0.985 & 0.993  \\

 &    &  & 20.0 &   &    0.923 & 0.940  \\

 &    &  & 30.0 &   &    0.885 & 0.909  \\

\hline
\multirow{3}{*}{5}    & \multirow{3}{*}{Si K} & \multirow{3}{*}{1845.0} 
           & 10.0 &  \multirow{3}{*}{sil.} &
                                          0.990 & 1.021  \\

 &    &  & 20.0 &   &    0.962 & 0.993  \\

 &    &  & 30.0 &   &    0.939 & 0.971  \\

\hline
\multirow{3}{*}{6}    & \multirow{3}{*}{Fe K} & \multirow{3}{*}{7123.0} 
           & 10.0 &  \multirow{3}{*}{sil.} &
                                          1.004 & 1.066  \\

 &    &  & 20.0 &   &    0.995 & 1.053  \\

 &    &  & 30.0 &   &    0.986 & 1.041  \\

\end{tabular}\\
\end{center}
$^a$ $j$ identifies the absorption edge.\\
$^b$ $E_{\rm edge}=\frac{1}{2}(E_{\rm max} + E_{\rm min})$ and $\Delta E=\frac{1}{2}(E_{\rm max} - E_{\rm min})$. $E_{\rm min}$ and $E_{\rm max}$ are the minimum and maximum energies, respectively, over which the fit was performed.\\
$^c$ $i$ identifies the material: 1=carbonaceous, 2=astrosilicate.\\
$^d$ \cite{Mathis+Rumpl+Nordsieck_1977}\\
$^e$ \cite{Weingartner+Draine_2001a}\\

\end{table}

\begin{figure}[!ht]
\centering
\includegraphics[width=0.8\linewidth]{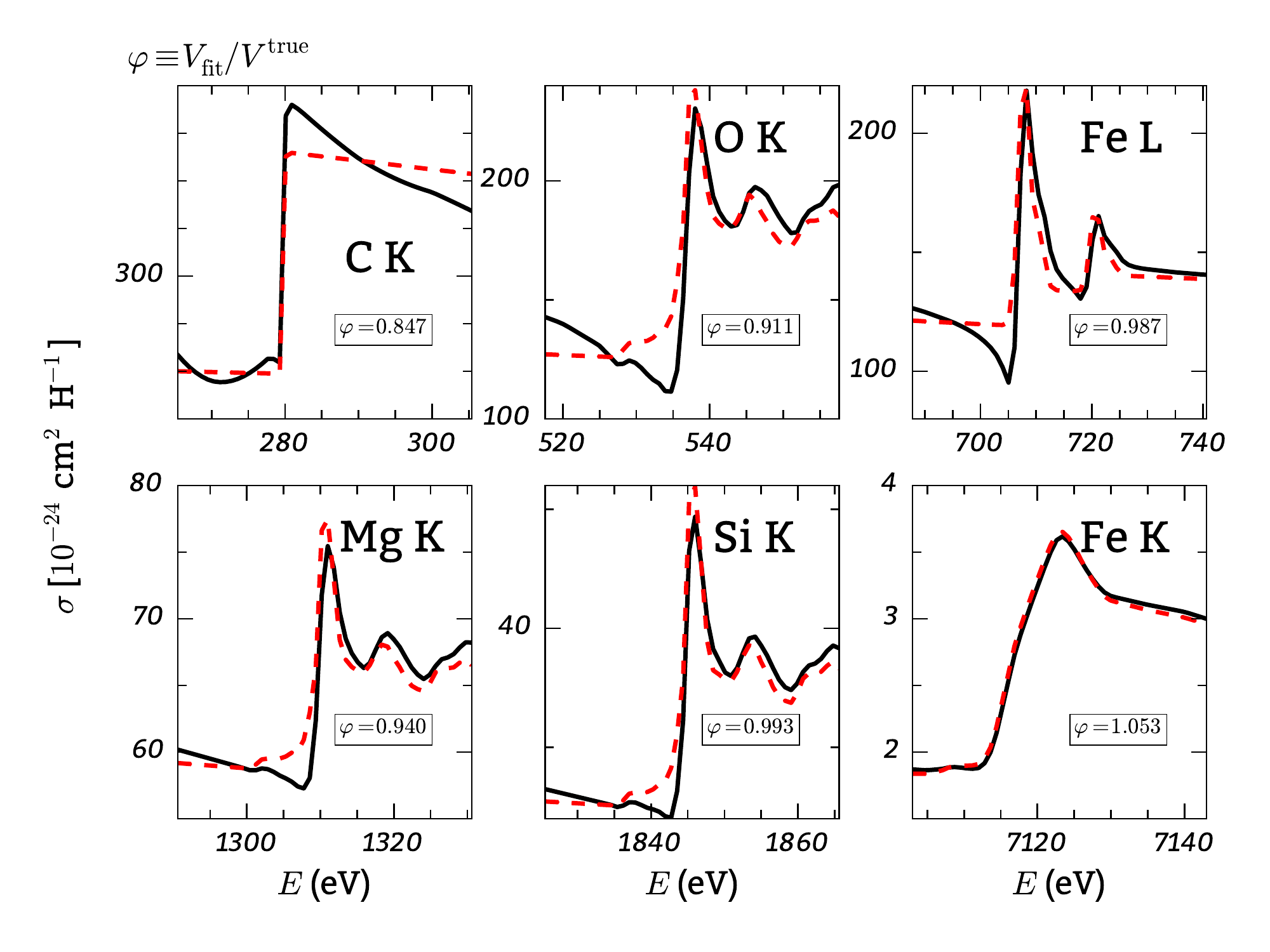}
\caption{\footnotesize\label{fig:fitabstoext_wd01} Fitting optical
  depth measurements without accounting for scattering by large dust
  grains produces errors in the inferred abundance constraints. The
  black solid line is the simulated (noise-free) optical depth for the
  WD01 dust size distribution of spherical grains for
  $R_V=3.1$. Volumes of silicate and carbonaceous materials were fit
  to all absorption edges individually. The red dashed line is the
  best fit for an optical depth model similar to
  \cite{Wilms+Allen+McCray_2000}, which ignores scattering.
  Abundance estimates are reasonably accurate (except for Carbon),
  but ignoring contributions from scattering significantly alters the
  absorption edge fine structure (except for Fe K).}
\end{figure}
\begin{figure}[!ht]
\centering
\includegraphics[width=0.8\linewidth]{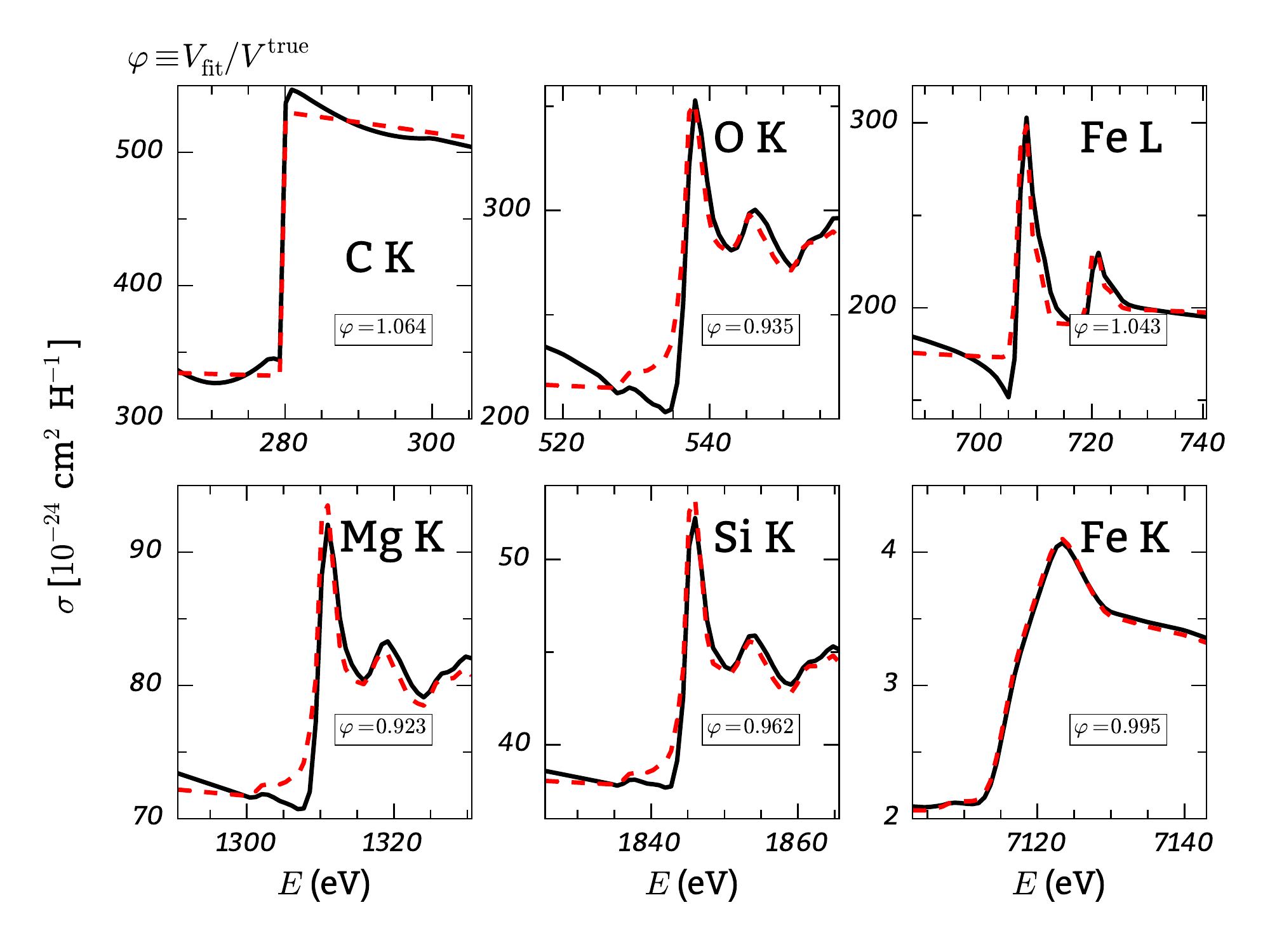}
\caption{\footnotesize\label{fig:fitabstoext_mrn} Similar to Figure
  \ref{fig:fitabstoext_wd01}, only the underlying size distribution
  was changed from \cite{Weingartner+Draine_2001a} to an MRN size
  distribution.}
\end{figure}

The fits (using Eq. \ref{eq:sigfit}) of the WAM2000-like attenuation model to six X-ray absorption
edges for the WD01 size distribution are shown in Figure
\ref{fig:fitabstoext_wd01}, and for the MRN size distribution in
Figure \ref{fig:fitabstoext_mrn}.  Table \ref{fit-table} shows the
ratio of the grain mass estimated by the WAM2000-like fit to the
``true'' grain mass for various size distribution assumptions and
energy ranges.

We find that the WAM2000 attenuation model is able to provide
moderately accurate estimates of the elemental abundances for Fe L, Mg
K, Si K, and Fe K --- e.g. errors of $\lesssim 5\%$ for the abundance
of Fe based on the Fe L$_1$, L$_{2,3}$ edges. However, because the
WAM2000 model neglects scattering, the wavelength dependence of the
extinction is not well-modeled. Consequently, attempts to identify the
chemical state (\newedit{e.g.,} Fe metal \newedit{vs.} Fe$_3$O$_4$) from the details of the
edge profile are prone to error, as is evident in Figures
\ref{fig:fitabstoext_wd01} and \ref{fig:fitabstoext_mrn} from the poor
fits of the WAM2000 model to the ``true'' extinction profiles
calculated for the same material.

\section{General Geometry Anomalous Diffraction Theory (GGADT)
		\label{sec:ggadt}}

As demonstrated in Section \ref{sec:modeling:misuse}, the WAM2000
attenuation model --- which assumes pure absorption --- has systematic 
errors $\gtrsim10\%$ for energies below the Si K edge. A more accurate 
attenuation model is necessary to obtain reliable \newedit{abundances} 
from X-ray attenuation measurements. \newedit{Accurate modeling of attenuation
requires} an accurate complex 
dielectric function and an algorithm for computing absorption and 
scattering by the dust grains. ADT is a natural choice for \newedit{the latter}: 
it can handle non-spherical grain geometries, \newedit{is} accurate for 
large grains, and is \newedit{more computationally efficient than Mie
theory or Rayleigh-Gans}.

\subsection{Intuition behind ADT
	\label{sec:ggadt:intuition}}


ADT was invented by \cite{van_de_Hulst_1957} to treat the problem of
scattering and absorption by a particle that is optically soft ($|m-1|
\ll 1$) but large compared to the wavelength of incident light
($x\equiv ka \gg 1$).
\refereenote{Optically ``soft'' is the appropriate definition here;
optically ``thin'' refers to the case when $\Im{m}ka\ll 1$.}

Because $x\gg 1$, the concept of independent rays of light passing
through the grain is a valid approximation. And, because $|m-1|\ll1$,
refraction and reflection effects are small and may be ignored. The
grain can thus be thought of as providing local phase shifts to the
incident plane wave. Absorption of the incident plane wave also occurs
if $\Im{m}\neq0$.

Consider a grain of arbitrary geometry. Define a plane $S$ just beyond the
extent of the grain and normal to $\hat{z}$, the direction of
propagation. The plane, $S$, is located at $z=0$, and the grain is confined to
$z<0$. 
Define $\hat{x}$ and $\hat{y}$ to be orthogonal unit
vectors that lie in $S$ with both $\hat{x}$ and $\hat{y}$ orthogonal
to $\hat{z}$. Define the ``shadow function'' on $S$:
\begin{equation}
\label{shadowfunction}
f(x,y) = 1 - e^{i\Phi(x,y)},
\end{equation}
where $\Phi(x,y)$ is the complex phase shift:
\begin{equation}
\label{phasefunction}
\Phi(x,y) \equiv k\int_{-\infty}^0 \left[m(x,y,z)-1\right]dz,
\end{equation}
$m$ is the (complex) index of refraction, and $k\equiv 2\pi/\lambda$.

To obtain far-field solutions to the scattered field, Huygen's
principle is applied to the shadow function $f$ on surface $S$. We
refer the reader to \cite{Draine+Allaf-Akbari_2006} for a brief
derivation of the extinction, absorption, scattering, and differential
scattering cross sections in the context of ADT, and to
\cite{van_de_Hulst_1957} for a more rigorous and detailed
treatment. 

\noindent The ADT formulae for absorption, scattering, and extinction cross sections \newedit{are}:
\begin{eqnarray}
C_{\rm abs}  &=& \int_S\left(1-e^{-2\Phi_2}\right)dS \label{adt_sigmas_abs}\\
C_{\rm sca} &=& \int_S |f|^2 dS = \int_S (1 - 2\cos{\Phi_1}e^{-\Phi_2} + e^{-2\Phi_2}) dS\label{adt_sigmas_sca}\\
C_{\rm ext}  &=& 2\int_S \left(1 - \cos{\Phi_1}e^{-\Phi_2}\right) dS\label{adt_sigmas_ext},
\end{eqnarray}
\noindent where $\Phi_1 \equiv \Re{\Phi}$ and $\Phi_2 \equiv \Im{\Phi}$.
The differential scattering cross section is
given by
\begin{equation}
\frac{dC_{\rm sca}}{d\Omega} = \frac{|S(\mathbf{\hat{n}})|^2}{k^2},
\label{eq:dscadom}
\end{equation}
where
\begin{equation}
S(\nhat) = \frac{k^2}{2\pi} \int_S e^{ik(\nhat\cdot\vec{x})}f(\vec{x}){\rm d}\vec{S}.
\label{eq:snhat}
\end{equation}

\subsubsection{ADT for Spheres}
For spherical grains, ADT yields analytic expressions for $C_{\rm
  ext}$, $C_{\rm abs}$, and $C_{\rm sca}$:\footnote{The expression for
  $Q_{\rm ext}$ given in \cite{Draine+Allaf-Akbari_2006} Eq (16)
  contained a sign error.}
\begin{eqnarray}
\label{eq:Q_{ext} for sphere}
Q_{\rm ext} \equiv \frac{C_{\rm ext}}{\pi a^2} &=& 2 + \frac{4}{|\rho|^2}\left\{\cos2\beta - e^{-\rho_2}\left[\cos(\rho_1 - 2\beta) + |\rho|\sin(\rho_1 - \beta)\right]\right\}\\
\label{eq:Q_{abs} for sphere}
Q_{\rm abs} \equiv \frac{C_{\rm abs}}{\pi a^2} &=& 1 + \frac{e^{-2\rho_2}}{\rho_2^2} + \frac{e^{-2\rho_2} - 1}{2\rho_2^2}.\\
\label{eq:Q_{sca} for sphere}
Q_{\rm sca} \equiv \frac{C_{\rm sca}}{\pi a^2} &=& Q_{\rm ext}-Q_{\rm abs} .
\end{eqnarray}
where 
\begin{equation}
\rho\equiv 2ka(m-1),
\end{equation}
$\rho_1\equiv {\rm Re}(\rho)$, $\rho_2\equiv {\rm Im}(\rho)$, and $\beta \equiv \arccos{(\rho_1/|\rho|)}$.
$S(\theta)$ becomes\footnote{\cite{Draine+Allaf-Akbari_2006}
  contained two typographical errors in their Eq (14). Their
  $J_0(ka\theta\cos{u})$ should be replaced with $J_0(ka\sin\theta\cos
  u)$, and $e^{-i\rho\sin u}$ should be replaced with $e^{i\rho\sin
    u}$.}$^,$\footnote{A FORTRAN subroutine \texttt{adt.f} (not to be confused with GGADT) 
    to calculate $dC_{\rm sca}/d\Omega$ for spheres using ADT is available from
    \url{www.astro.princeton.edu/~draine/scattering.html}.}
\begin{equation}
S_{\rm ADT}(\theta) = (ka)^2\int_{0}^{\pi/2}du J_0(ka\sin\theta\cos u)\left(1 - e^{i\rho\sin u}\right)\sin u \cos u,
\end{equation}
where $J_0$ is the Bessel function of the first kind (of order
zero); for $\theta=0$, 
\begin{equation}
S_{\rm ADT}(\theta=0) = \frac{1}{2} + \frac{1 + ie^{i\rho}(\rho + i)}{\rho^2}.
\end{equation}

\subsubsection{ADT Accuracy}
ADT provides a natural complement to the Rayleigh-Gans approximation
in the X-ray regime; small grains are accurately modeled by the
Rayleigh-Gans approximation (or Mie theory), while larger grains are
accurately modeled by ADT. There is an analytic solution to the ADT
equations for spherical scatterers, so the computational cost of using
ADT over Rayleigh-Gans is negligible. It is also reasonably
straightforward to extend ADT to other geometries.

\begin{figure}[!ht]
\centering
\includegraphics[width=0.45\linewidth]{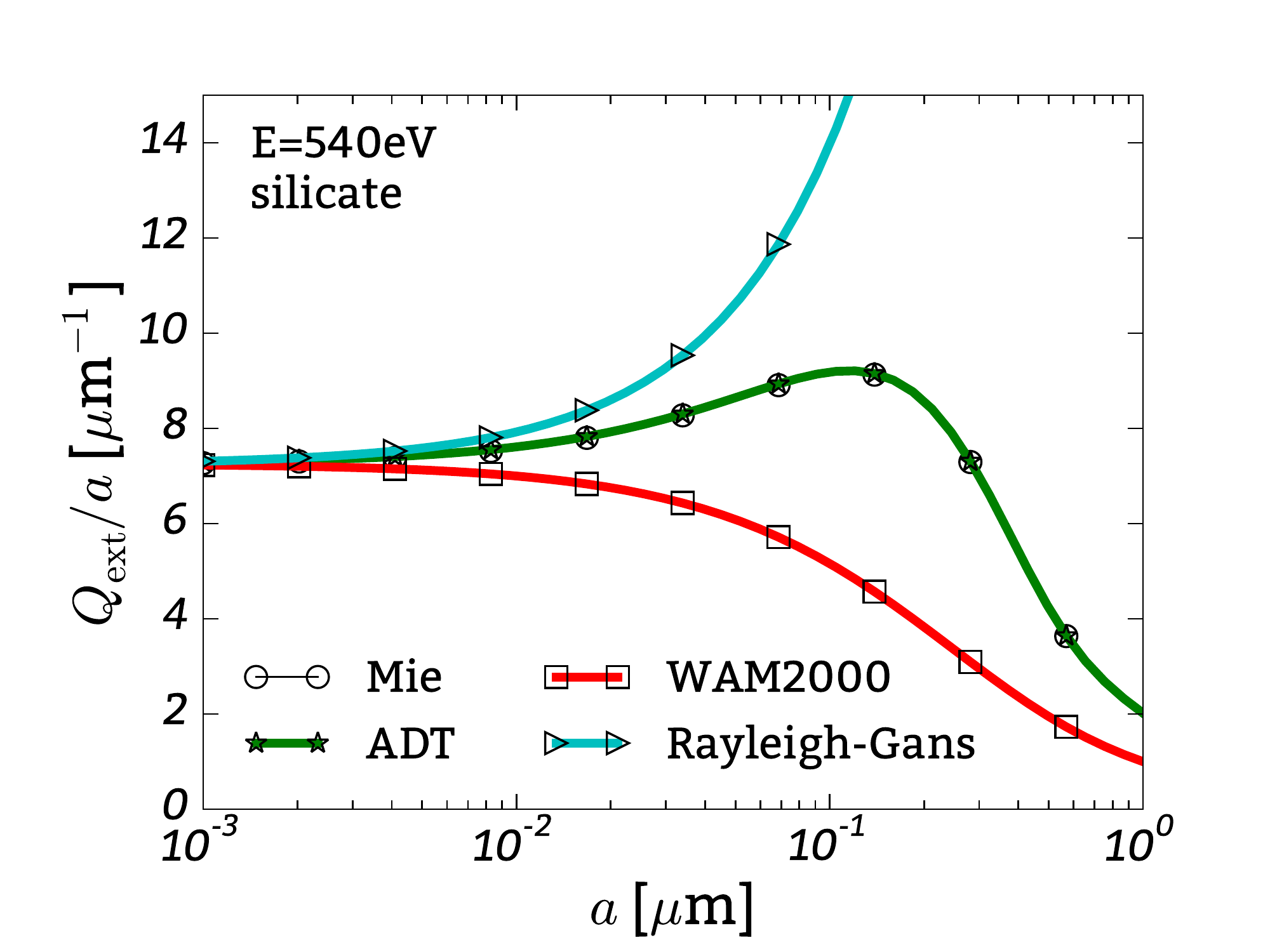}
\includegraphics[width=0.45\linewidth]{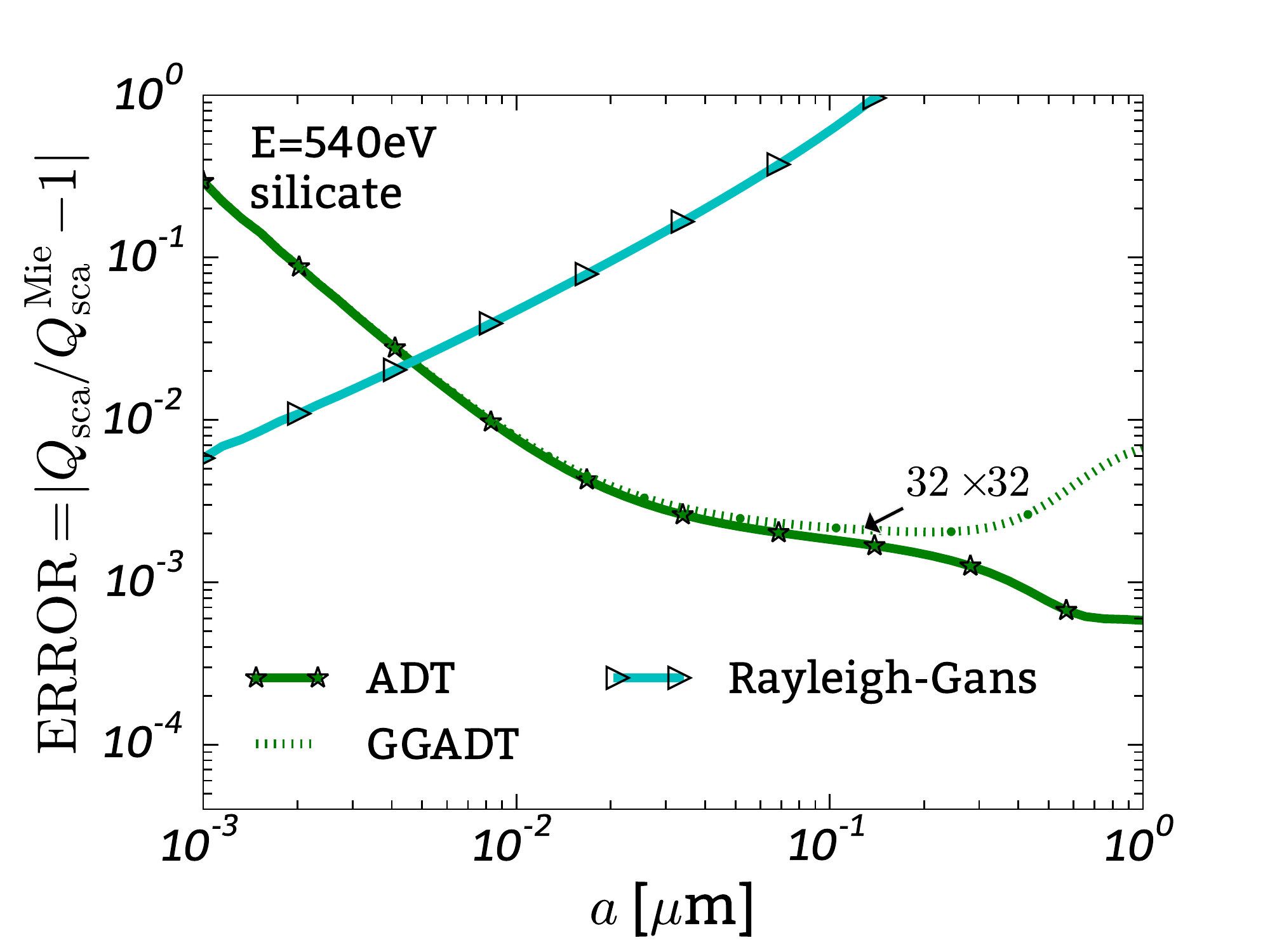}
\includegraphics[width=0.45\linewidth]{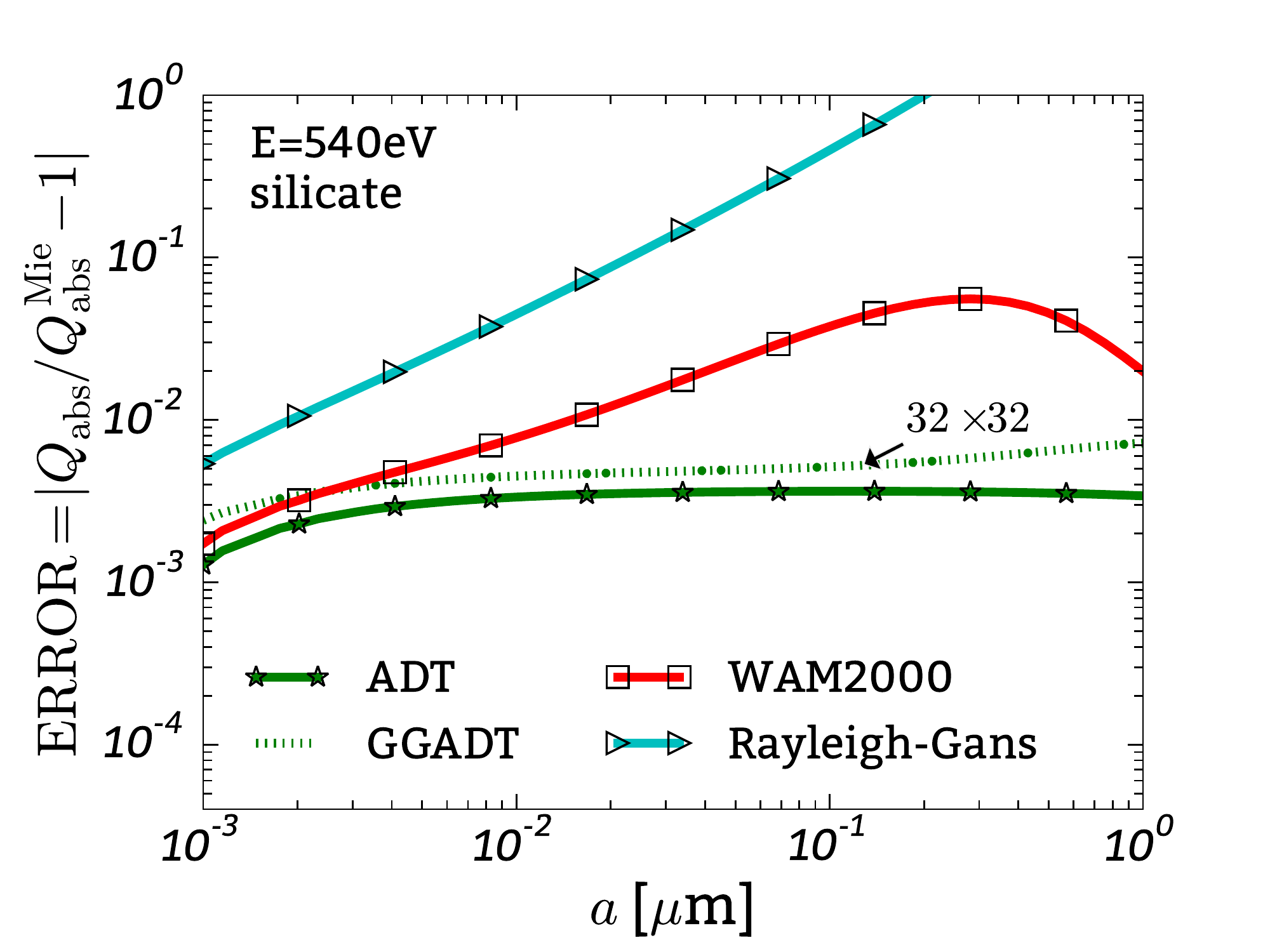}
\includegraphics[width=0.45\linewidth]{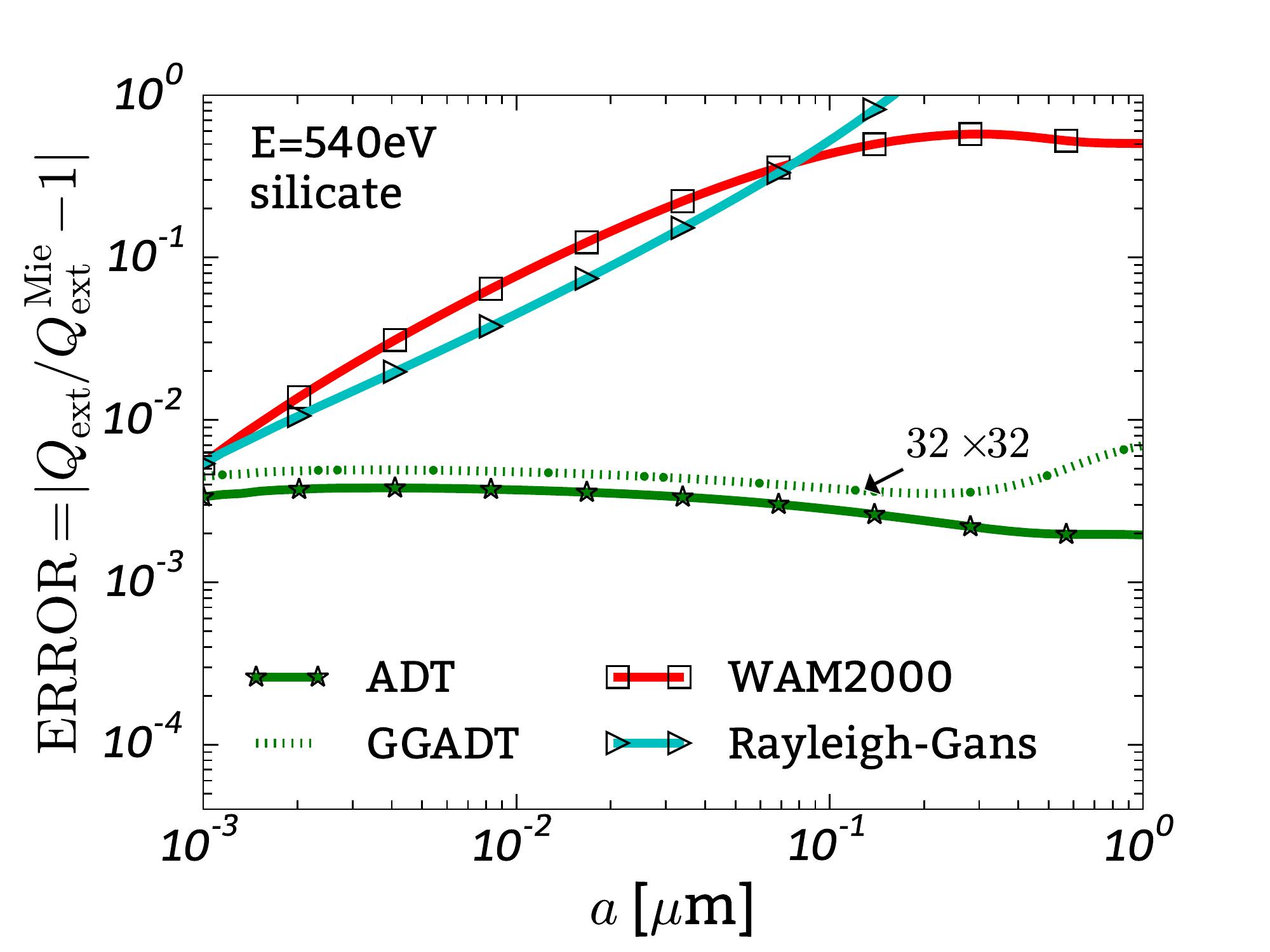}
\caption{\footnotesize\label{fig:qext_vs_a} {\it Top left}: Comparing
  calculated $Q_{\rm ext}$ at $h\nu=540$ eV (O K edge) to exact
(Mie theory) result.
  ADT calculations of $Q_{\rm ext}$ are
  accurate even for small grains. {\it Top right, bottom left, and
    bottom right}: fractional error of $Q_{\rm sca}$, $Q_{\rm abs}$,
  and $Q_{\rm ext}$, respectively.
  At small grain sizes, the ray optics approximations
  made in ADT fail and ADT becomes unsuitable for estimating the
  scattering cross section; however, in this regime, scattering is in
  any case negligible. The error in $Q_{\rm ext}$ calculated with ADT
  remains below 0.5\% for all grain sizes. The short-dashed line shows
  numerical ADT calculations performed with GGADT using a coarse $32\times32$
  numerical grid to represent the shadow function. Even using a coarse $32\times32$ 
  grid, the accuracy of GGADT is still better than 1\%. }
\end{figure}

Even for small grains where the ``ray optics'' approach of ADT fails, 
ADT will still produce accurate estimates of extinction, since, in this case, 
the extinction is dominated by absorption. To illustrate this, 
Figure \ref{fig:qext_vs_a} (top left) compares $Q_{\rm abs}$, 
$Q_{\rm sca}$, and $Q_{\rm ext}$ for silicate spheres calculated with 
Mie theory, ADT, the Rayleigh-Gans approximation, and the WAM2000 
absorption-only estimate. 

\subsection{GGADT: General Geometry Anomalous Diffraction Theory
	\label{sec:ggadt:code}}

The authors have written a Fortran 95 program GGADT
that uses ADT to calculate (1)
the energy-dependent scattering and absorption cross sections, and (2)
the differential scattering cross section for grains of arbitrary
geometry and composition. GGADT is fast, portable, GNU-compliant, and
well documented.\footnote{GGADT can be downloaded from:
  \newedit{\url{http://www.ggadt.org}}.} GGADT uses the General
Prime Factor Algorithm (GPFA) of \cite{Temperton_1992} to do fast
Fourier transforms. A brief description of GGADT usage \newedit{can be found in
Appendices \ref{ggadt_desc} and \ref{examples_ggadt}}.

\subsubsection{An application of GGADT: effect of grain geometry on X-ray extinction}

The geometry of dust grains is not currently well
constrained. Polarization of starlight implies that dust grains are
not spherically symmetric. The next simplest grain geometry is the
spheroid; spheroidal grain models are able to reproduce
observations of starlight polarization and extinction
\citep{Kim+Martin_1995,Draine+Fraisse_2009}.

However, for plausible dust evolution scenarios, dust grains are
likely more complicated than single-material spheroids or even
ellipsoids. ISM grains could have irregular geometries as well as
inhomogeneous composition. Some authors
\citep[e.g.,][]{Mathis+Whiffen_1989,Henning+Stognienko_1993,Stognienko+Henning+Ossenkopf_1995}
have argued for highly porous geometries.

To illustrate the possible effects that grain geometry might have on
abundance measurements based on X-ray extinction, we employ GGADT to
compute the extinction cross sections for five example grain
geometries. The size is specified by the radius of an equal-volume
sphere, 

\begin{equation}
\label{eq:aeff}
\aeff\equiv(3V/4\pi)^{1/3}
\end{equation}

\noindent where $V$ is the volume of the
solid material. 
The five grain geometries used are (1) a sphere, (2)
an oblate spheroid, (3) a prolate spheroid, (4) a BAM2 aggregate, and
(5) a BA aggregate, each with the same mass ($\aeff = 0.2\mu$m) and
silicate composition.

\begin{figure}[t]
\centering
\includegraphics[width=0.45\linewidth,trim={0 5.0cm 0 3.5cm},clip]{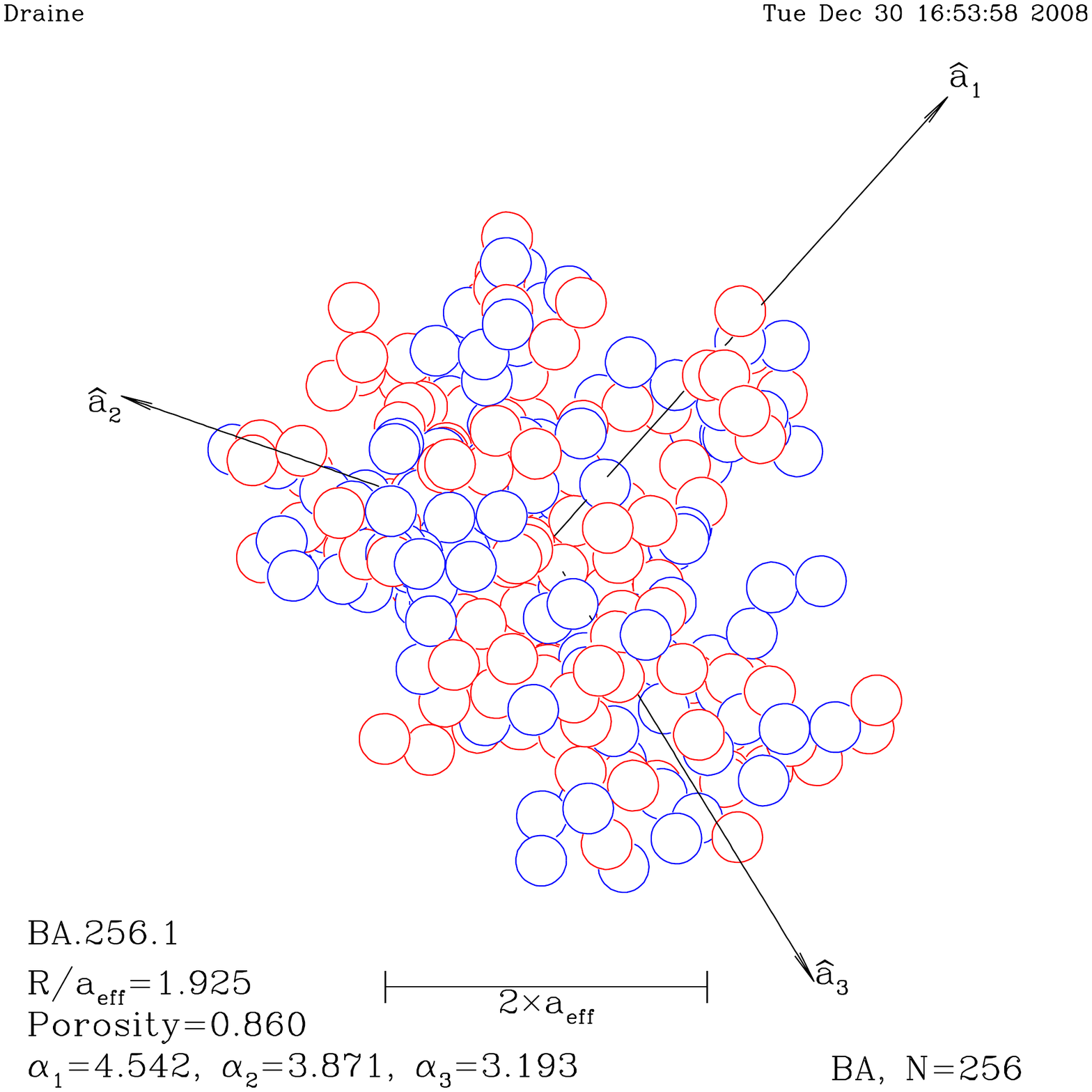}
\includegraphics[width=0.45\linewidth,trim={0 5.0cm 0 3.5cm},clip]{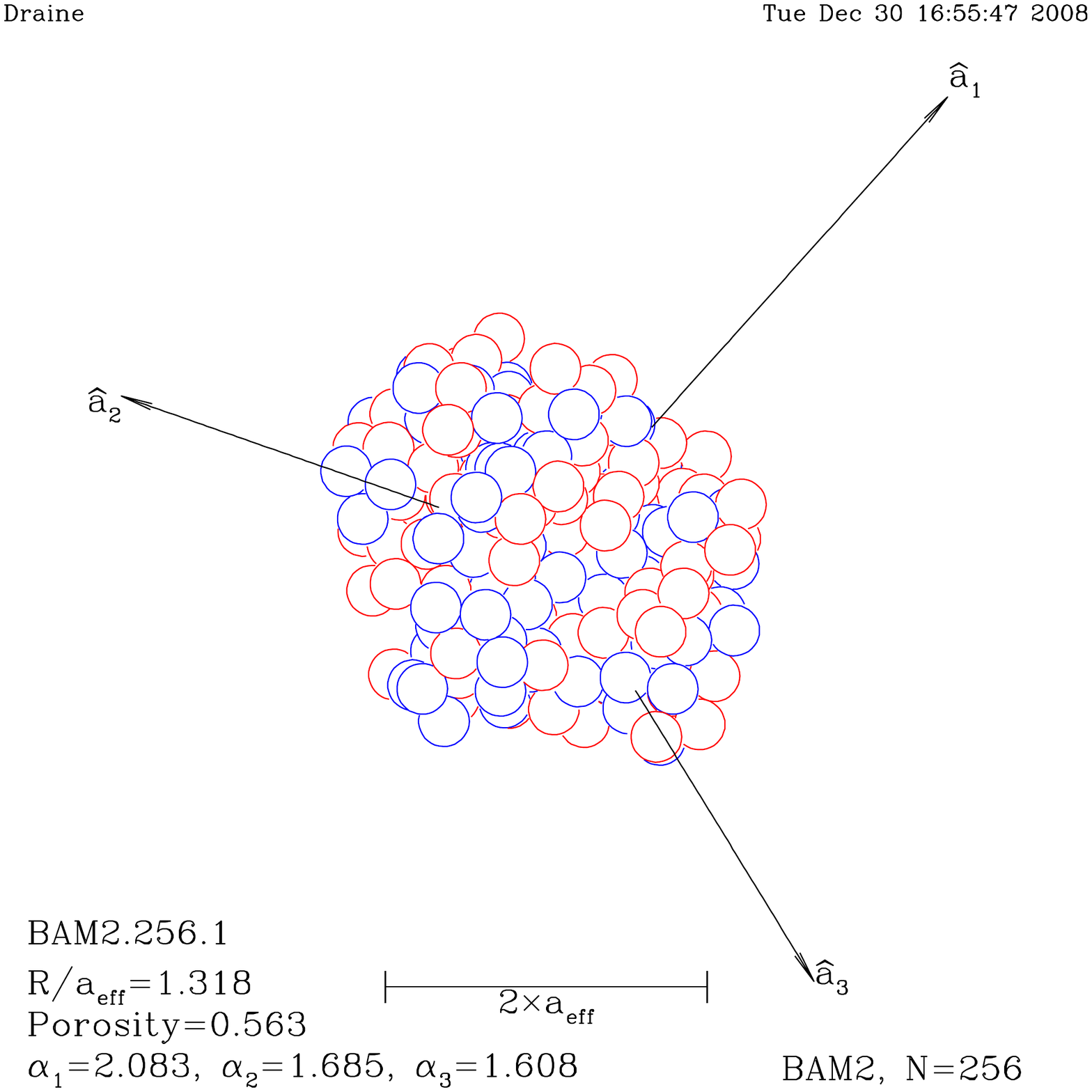}
\caption{\footnotesize\label{agglomgrains}Two random aggregates used to investigate
  the effects of grain geometry and porosity on dust extinction. Both
  figures are from \cite{Shen+Draine+Johnson_2008}. {\it Left}: a
  porous BA grain composed of $N=256$ monomers. {\it Right}: a less
  porous random aggregate produced by the BAM2 algorithm, also
  containing $N=256$ monomers.}
\end{figure}

Ballistic agglomeration (BA) aggregates are constructed by single-size
spherical monomers arriving on random trajectories and adhering to
their initial point of contact; BAM2 aggregates require
arriving monomers (after the third) to make contact with
a total of three other monomers prior to the arrival of the next monomer. BAM2 aggregates
have porosities $P$ significantly less than that of BA
aggregates. A detailed description of BA and BAM2 agglomeration 
is given in \cite{Shen+Draine+Johnson_2008}.
Figure \ref{agglomgrains} shows examples of BA and BAM2 agglomerates.

This paper employs the definition of porosity from
\cite{Shen+Draine+Johnson_2008}. For a given grain, define an
equivalent ellipsoid (EE) as the uniform density ellipsoid that has
the same mass and moment of inertia tensor as the grain. The porosity
of the grain $P$ is then defined by $\aee^3(1 - P) \equiv \aeff^3$
where $(4\pi/3)\aee^3$ is the volume of the equivalent
ellipsoid. Thus, $\aee = \aeff(1 - P)^{-1/3}$.

\begin{figure}[t]
\centering
\includegraphics[width=0.9\linewidth]{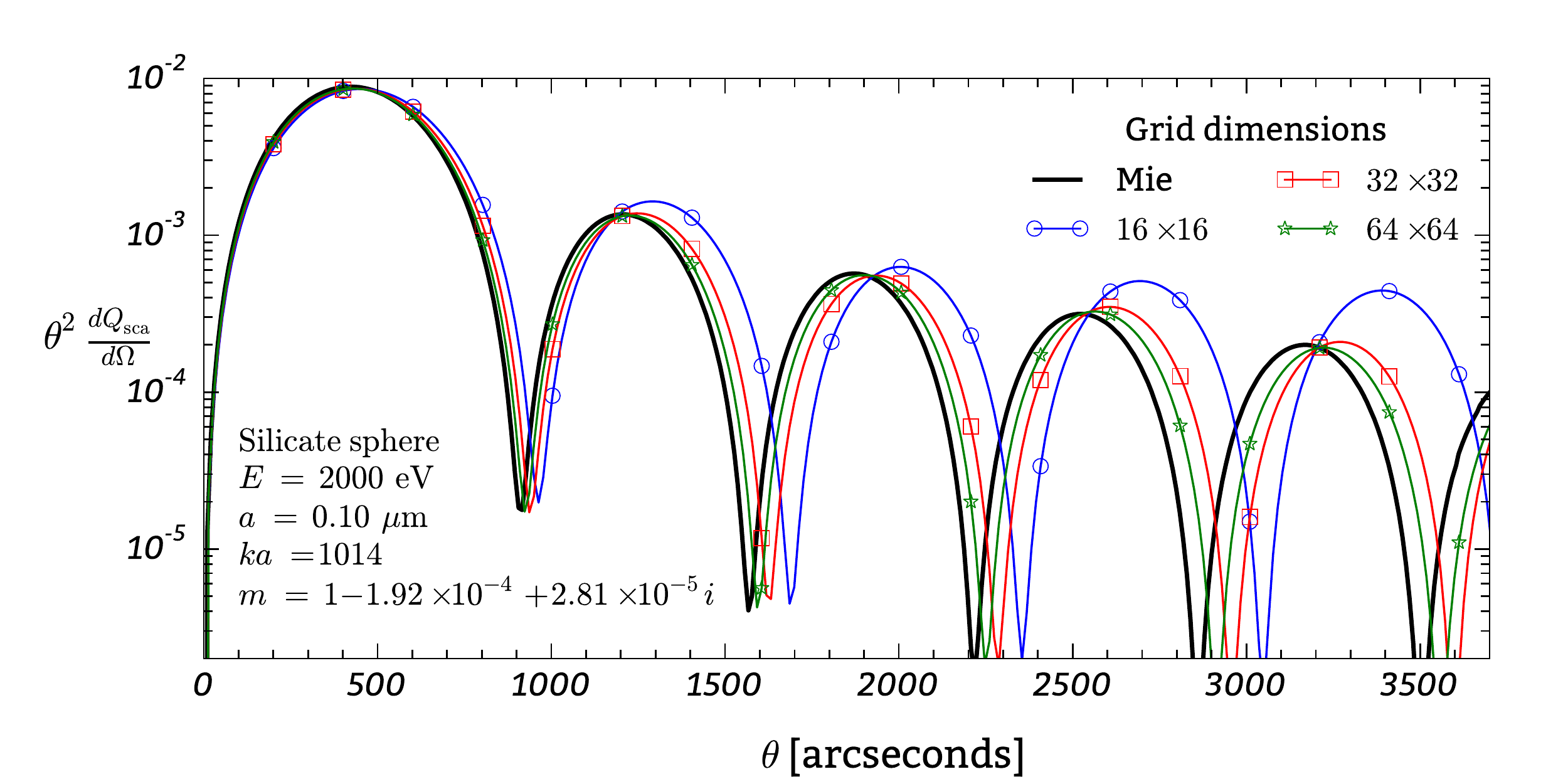}
\caption{\footnotesize\label{diffscat} GGADT results for spheres
  using different grid resolutions ($N_x\times N_y$), compared
  with the Mie theory result for spheres. $64\times64$ grids produce results with
  accuracies comparable to that of larger grids.}
\end{figure}

For grains with complex geometries, such as BA and BAM2 aggregates, \newedit{the} shadow function,
$f(x,y)$ (see Eq. \ref{shadowfunction}) is \newedit{evaluated}
on an $N_x\times N_y$ grid, \newedit{\sout{in order}} to facilitate the computation of discrete Fourier transforms.  
The choice of $N_x$ and $N_y$ determines how accurately
$f(x,y)$ will be described on small linear scales.
Figure \ref{diffscat} shows how the GGADT result for a sphere
depends on the chosen grid size.  We see that $64\times64$ yields
results that are accurate to a few percent (heights of scattering peaks, and
positions of maxima and minima). The reason that high-resolution numerical representations
(e.g. 2048$\times$2048) of the shadow function \newedit{are not} required to achieve ~1\% accuracy
arises from the nature of equation (\ref{eq:snhat}); for small-angle scattering ($\sin\theta \ll 1$),
only the long-wavelength contributions to the Fourier transform of $f$ are relevant to
the calculation of $d\sigma_{\rm sca}/d\Omega$.
We will generally use grid resolutions $128\times128$ or higher for calculations
in this paper unless otherwise specified.
  
\begin{figure}[t]
\centering
\includegraphics[width=0.9\linewidth]{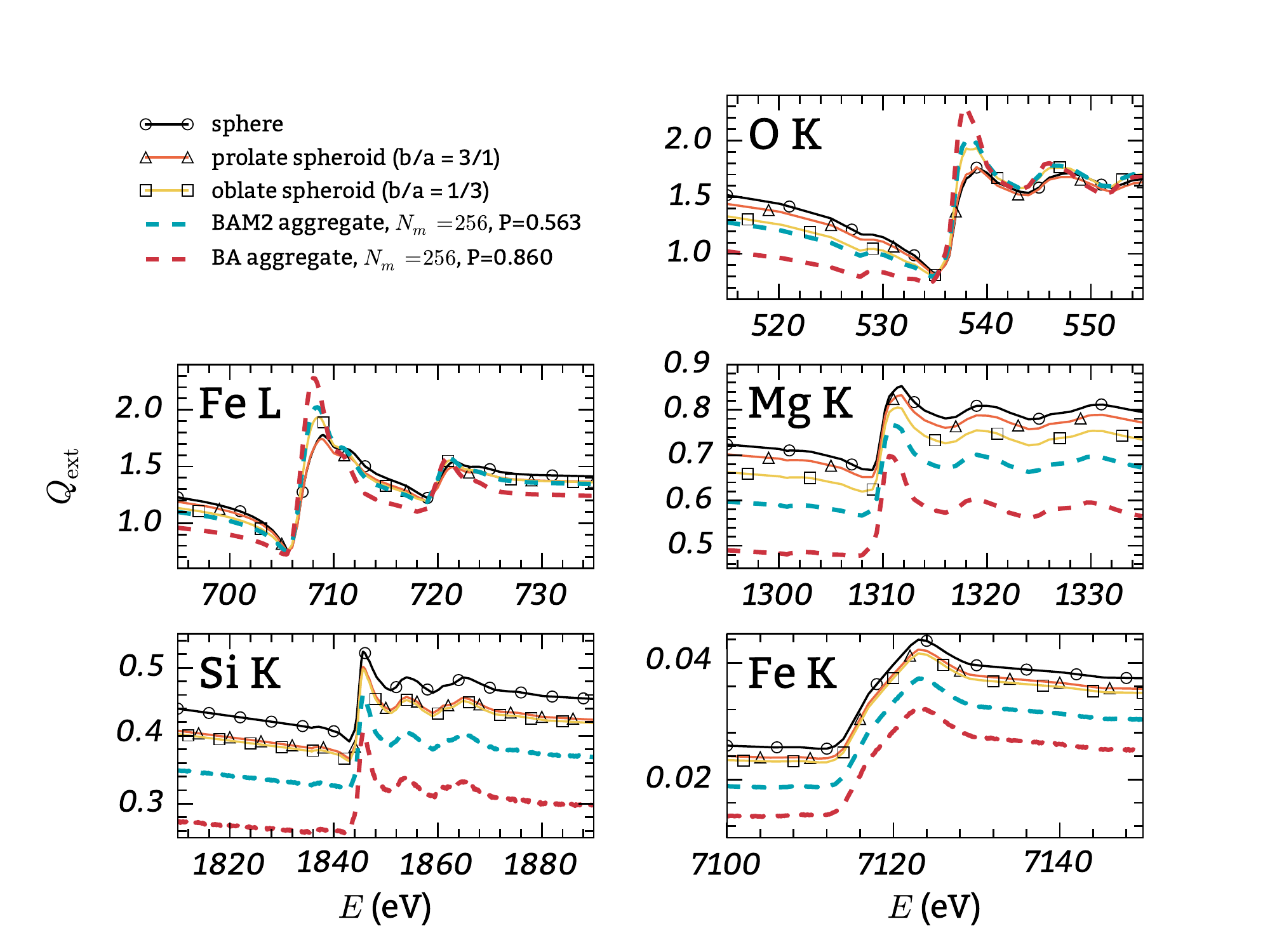}
\caption{\footnotesize \label{extvsgeom_abs} Orientation-averaged
  $Q_{\rm ext}$ for equal-mass $\aeff = 0.2\mu{\rm m}$ silicate
  grains with different geometries. A $256\times 256$ grid was used for
  the shadow function in all cases, and calculations were
  averaged over 64 random orientations. Porous, extended grain
  geometries significantly alter the fine structure of the absorption
  edges (except for the Fe K edge).
  Moderately prolate/oblate spheroidal grains, on the
  other hand, have $Q_{\rm ext}$ very similar to spherical grains.}
\end{figure}
\begin{figure}[ht]
\centering
\includegraphics[width=0.9\linewidth,clip=true,
                 trim=0.0cm 0.0cm 0.0cm 1.0cm]{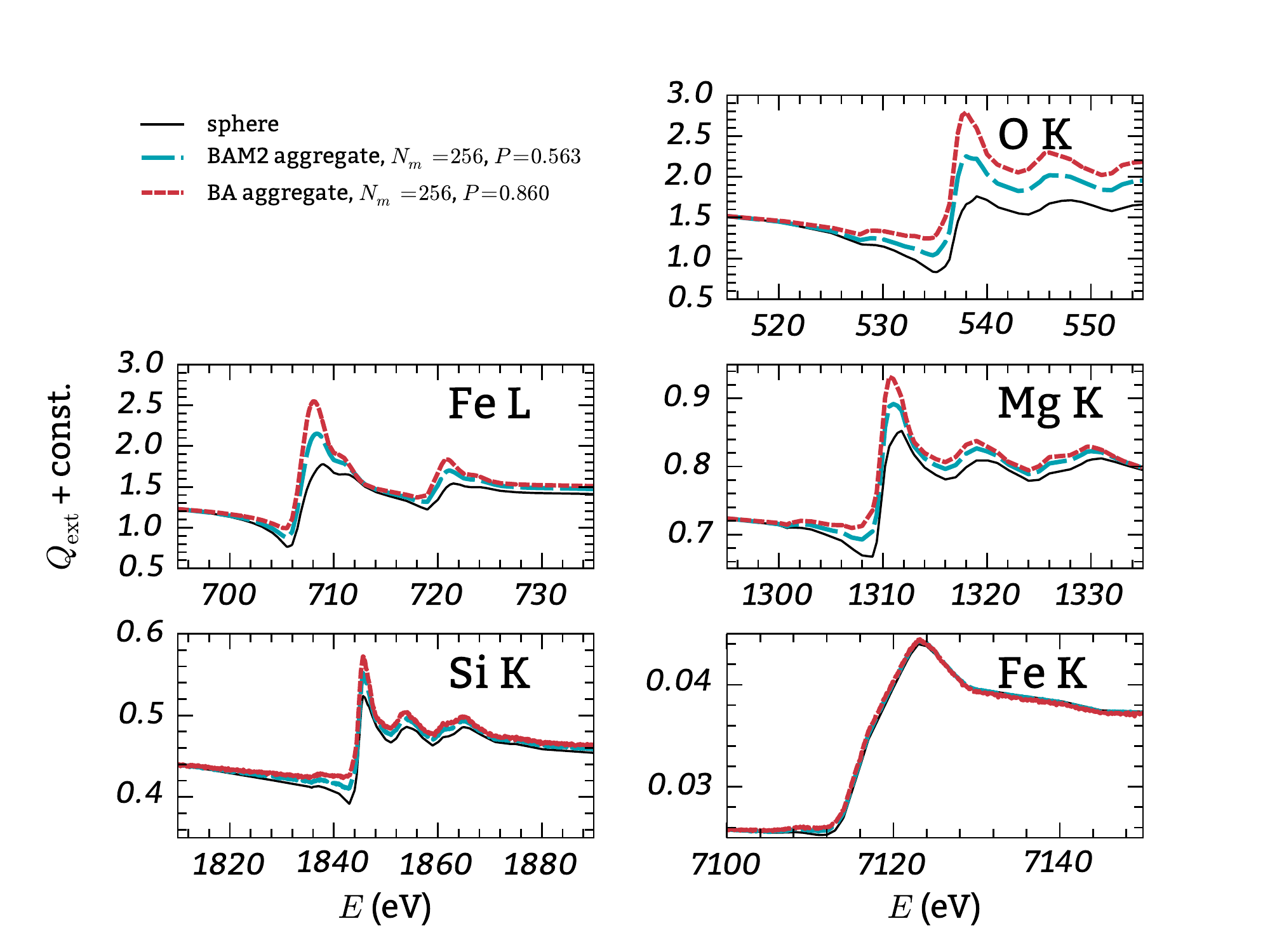}
\caption{\footnotesize\label{extvsgeom} Same as Figure
  \ref{extvsgeom_abs} (silicate, $a_{\rm eff}=0.2\mu$m), 
  but omitting spheroidal grains, and adding offsets so
  that all curves match $Q_{\rm ext}$ for a sphere
  at the left-most part of each plot.}
\end{figure}

Extinction cross sections calculated for several $\aeff=0.2\mu$m
grains are shown in Figures \ref{extvsgeom_abs} and
\ref{extvsgeom}. The dielectric function of the grains was taken to be
that of MgFeSiO$_4$ olivine described in \cite{Draine_2003c}. All
cross sections are averaged over 64 random grain orientations. All
calculations were done using ADT. For the spherical grain,\newedit{\sout{ calculated}}
results \newedit{calculated with GGADT} are indistinguishable from Mie theory. For the spheroidal and
agglomerate grains, the extinction was calculated using the GGADT
code, with the shadow function $f(x,y)$ (see Equation
\ref{shadowfunction}) sampled on a 512$\times$512 grid, providing
excellent accuracy (see Figure \ref{diffscat} below). The aggregates
used in Figures \ref{extvsgeom_abs} and \ref{extvsgeom} are those in
Figure \ref{agglomgrains}.

\subsection{Porosity}

\begin{figure}[!ht]
\begin{center}
\includegraphics[width=0.9\linewidth]%
{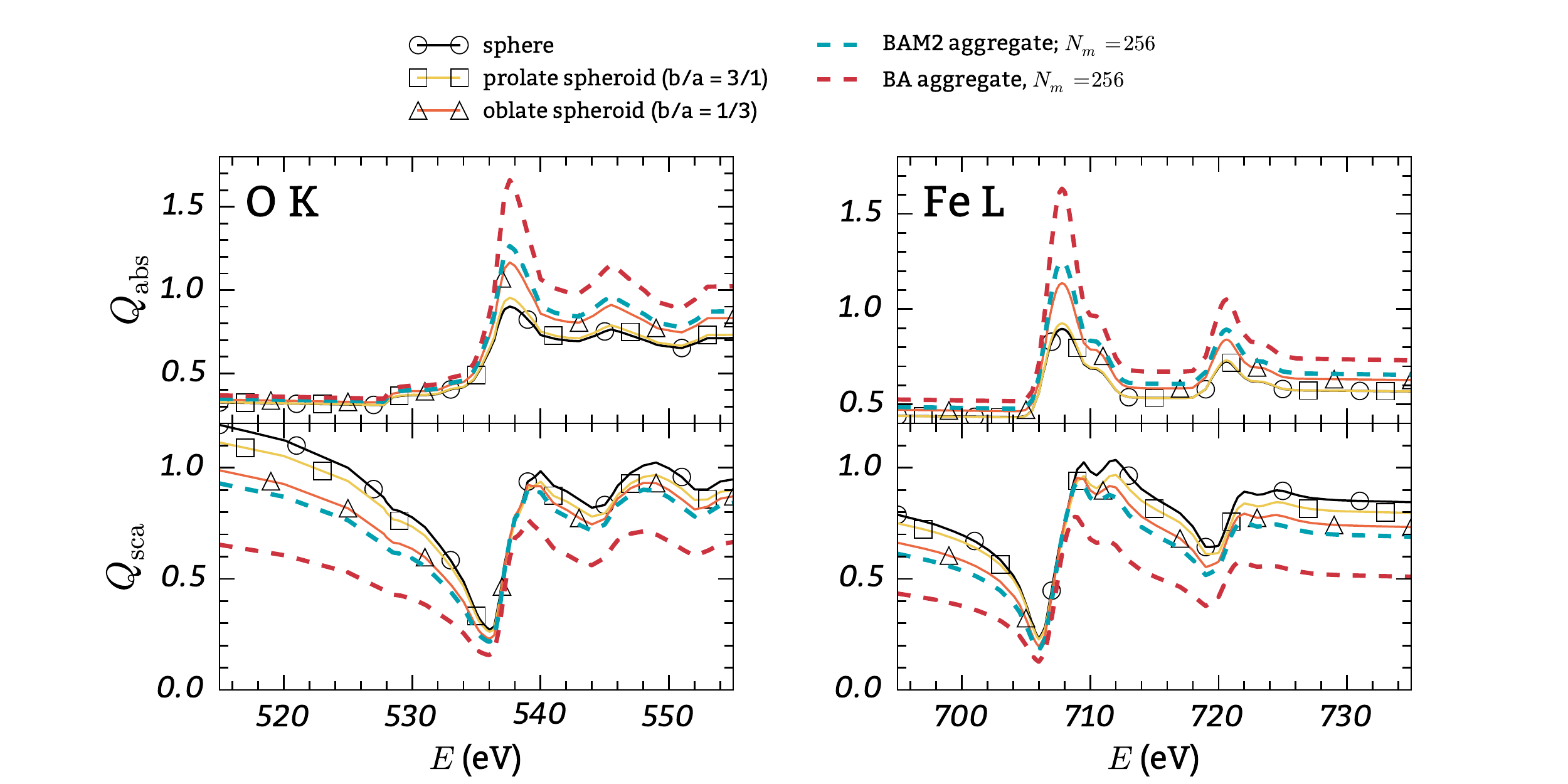}
\end{center}
\caption{\footnotesize\label{absscatvsgeom} Effect of grain
  geometry on absorption and scattering for $a_{\rm eff}=0.2\micron$
  silicate grains, near the O K and Fe L
  edges.}
\end{figure}

As stated earlier in this section, higher porosity increases
absorption efficiency. This trend, along with the effect of porosity
on scattering, is shown in Figure \ref{absscatvsgeom}. For
$\aeff=0.2\mu$m grains, the scattering efficiency decreases as
porosity increases. The extinction efficiency decreases with increased
porosity except near the Fe L and O K absorption edges, where the
increase in absorption efficiency dominates over the decreasing
scattering efficiency.

Orientation-averaged X-ray extinction cross sections do not differ
appreciably between the spherical and spheroidal
dust grains in Figure \ref{extvsgeom_abs}. However, the absorption and scattering
cross sections of random aggregates, as shown in Figure
\ref{absscatvsgeom}, do deviate significantly from those of
spherical grains. The absorption efficiency of random aggregates is
increased at all photon energies relative to an equal-mass sphere;
volume elements of porous grains are exposed to a larger fraction of
the incident light than compact grains, and therefore the grain as a
whole absorbs light more efficiently. In compact grains, parts of the
grain are ``shadowed'' (i.e., exposed to less of the incident light)
by grain material closer to the light source, and thus these
shadowed regions do not absorb as much light as in porous grains.

The effects of porosity on the scattering cross section are more
complicated, and we pause to provide a semi-analytical discussion of
two limiting cases (the optically thin and optically thick regimes)
below.

In the regime where $|\Phi|\ll 1$ (the optically thin regime),
Equation (\ref{adt_sigmas_sca}) for spherical grains becomes
\begin{eqnarray}
C_{\rm sca} &=& \int|1 - e^{i\Phi}|^2dxdy \\
			 &\approx& \int|1 - (1 + i\Phi)|^2dxdy \\
			 &=& \int|\Phi|^2dxdy\\
			 &=& k^2|m_{\rm eff}-1|^2\int_0^{2\pi}\int_0^{\aee}Z(r)^2 rdrd\theta,
\end{eqnarray}
where $Z(r) \equiv 2\sqrt{\aee^2 - r^2}$ is the length of the grain along the $z$ axis at $r = \sqrt{x^2 + y^2}$, and $m_{\rm eff}$ is the effective index of refraction which depends upon the porosity of the grain,
\begin{equation}
(m_{\rm eff} -1) \approx (m_0 - 1)(1 - P),
\end{equation} 
where $m_0$ is the index of refraction for the grain material.

Thus, in the optically thin regime,
\begin{eqnarray}
C_{\rm sca} &\approx& 2\pi k^2\left(\aeff^4(1-P)^{-4/3}\right)\left(|m_0 - 1|^2(1-P)^2\right)\\
			&=& 2\pi k^2 \aeff^4|m_0 - 1|^2(1-P)^{2/3}
\end{eqnarray}
\noindent which implies that an increase in porosity produces a
\emph{decrease} in the scattering cross section. This accounts for the
decrease in scattering efficiency at the Fe L edge shown in Figure
\ref{absscatvsgeom}. However, lower energy absorption edges (e.g. the
O K edge), exhibit smaller decrements in the scattering cross section
close to the absorption edge. This is because, as will be discussed
below, porosity has the opposite effect on scattering when not in the
optically thin regime.

In the opaque limit [$(ka){\rm Im}(m-1)\gg 1$], however, we have 
that $C_{\rm sca}\approx C_{\rm abs}\approx \Aproj$ so $C_{\rm ext} \approx
2\Aproj$ (see \cite{Bohren+Huffman_1983} for discussion of the
``extinction paradox''), where $\Aproj$ is the projected area of the
grain along the direction of propagation.

For fixed $\aeff$, increased porosity $P$ leads to a
\emph{larger} $\Aproj$. Thus,
in the opaque limit, increasing the porosity of a given grain
will cause an \emph{increase} in $C_{\rm sca}$.

As can be seen in Figures \ref{extvsgeom_abs} and \ref{extvsgeom} the
absorption edge fine structure is significantly affected by grain
geometry (especially porosity). Sensitivity to dust geometry means
that absorption-edge fine structure is indeed a laboratory for
constraining dust geometry as well as composition, but also calls into
question the significance of claimed abundance constraints derived
from fits to X-ray absorption edge fine structure.

 \subsection{Effective medium theory calculations}

 \begin{figure}[ht]
\centering
\includegraphics[width=0.9\linewidth]{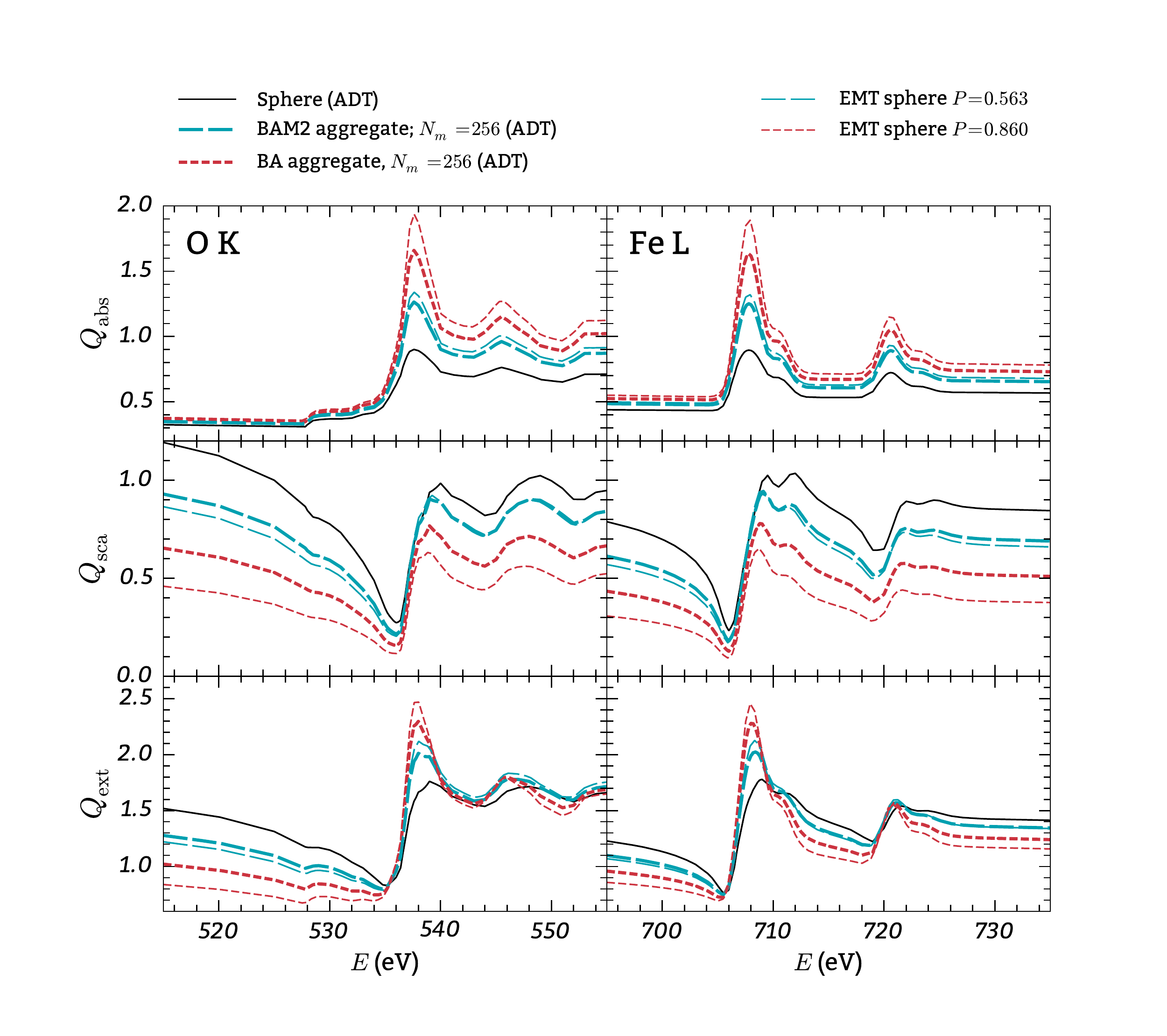}
\caption{\footnotesize\label{fig:emt_results} Comparing EMT and ADT 
calculations of absolute absorption and scattering cross sections for 
$\aeff=0.2\mu{\rm m}$ grains. EMT assumes a uniform density within a 
sphere of radius $\aeff(1 - P)^{-1/3}$, while the ADT calculations are done
for aggregates of 256 identical spherical monomers.
The EMT calculations are done for spheres with the same porosity $P$
as the corresponding aggregate grain.}
\end{figure}

Effective medium theories (EMT) have been used since \cite{Maxwell-Garnett_1904} and
\cite{Bruggeman_1935} to obtain an effective dielectric 
function, $\epsilon_{\rm eff}$ for an inhomogeneous particle composed of several
materials with dielectric functions $\epsilon_1$, $\epsilon_2$, $\dots{}$. 
If the particle itself is approximately spherical, Mie theory can then be used to
calculate scattering and absorption by the particle.

More sophisticated effective medium theories \newedit{\cite[see, e.g.,][]{Stognienko+Henning+Ossenkopf_1995}} have
been constructed to try to take into account the shapes of inclusions and voids in inhomogeneous grains
with complex structures. \cite{Valencic+Smith_2015} investigated how well observations 
of X-ray halos could be modeled by a variety of dust models,
including those of \cite{Zubko+Dwek+Arendt_2004} (hereafter ZDA2004), and concluded that the families of
porous dust models described in ZDA2004 did not fit observations as well as models without porosity.

At X-ray energies, the wavelength may be comparable or even smaller than the sizes of
inclusions and voids, and the validity of EMT is uncertain. Furthermore, EMT
aims to reproduce the mean polarization, while scattering depends on the mean square
polarization, and hence it is of interest to compare ADT calculations
against those of EMT to determine how well EMT approximates the scattering and absorption
properties of BA and BAM2 grains. 

Figure \ref{fig:emt_results} compares ADT calculations of scattering
and absorption by ballistic agglomerates against Mie theory calculations for a sphere
with radius $a = \aee$ and with a uniform index of refraction $m=m_{\rm eff}$ obtained
from EMT. At X-ray energies, the different EMTs [e.g., \cite{Maxwell-Garnett_1904} and
\cite{Bruggeman_1935}] give essentially identical $m_{\rm eff}$.
EMT agrees best with the most compact ballistic agglomerations
(BAM2), but systematically underestimates the scattering cross section and 
overestimates the absorption cross sections of the ballistic agglomerate grains. These
effects partially cancel when the extinction is computed, but the height of the absorption edge --- a key
diagnostic for ISM abundances --- remains overestimated. The accuracy of the EMT approach decreases
with increasing porosity.


\section{Differential Scattering Cross Section}

\begin{figure}[!ht]
\centering
\includegraphics[width=0.9\linewidth]{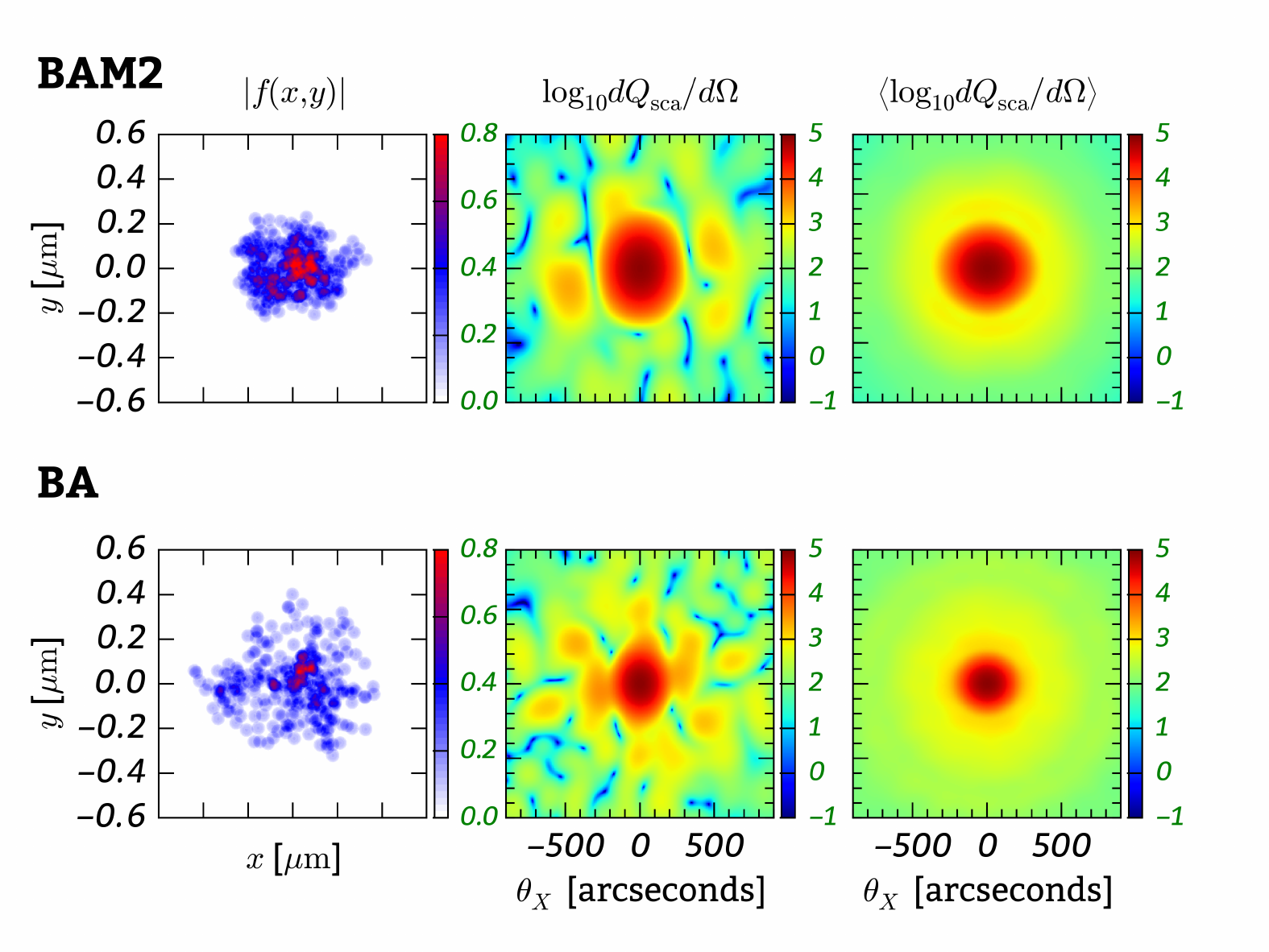}
\caption{\footnotesize\label{2dheatmaps} Output from GGADT for a
  BAM2 (\emph{top}) and a BA (\emph{bottom}) grain
  with 256 silicate monomers (both are shown in Figure \ref{agglomgrains}). For
  each grain, $a_{\rm eff}=0.2\mu{\rm m}$, and $512\times512$ grids are used to
  represent the shadow function. 
  \emph{Left:} the magnitude of the shadow function for a
  single orientation and $h\nu=2$keV. \emph{Middle:} $dQ_{\rm
    sca}/d\Omega$ for that same single orientation; \emph{Right:}
  $dQ_{\rm sca}/d\Omega$ averaged over 300 random (3D)
  orientations.}
\end{figure}
\begin{figure}[!ht]
\centering
\includegraphics[width=0.9\linewidth]{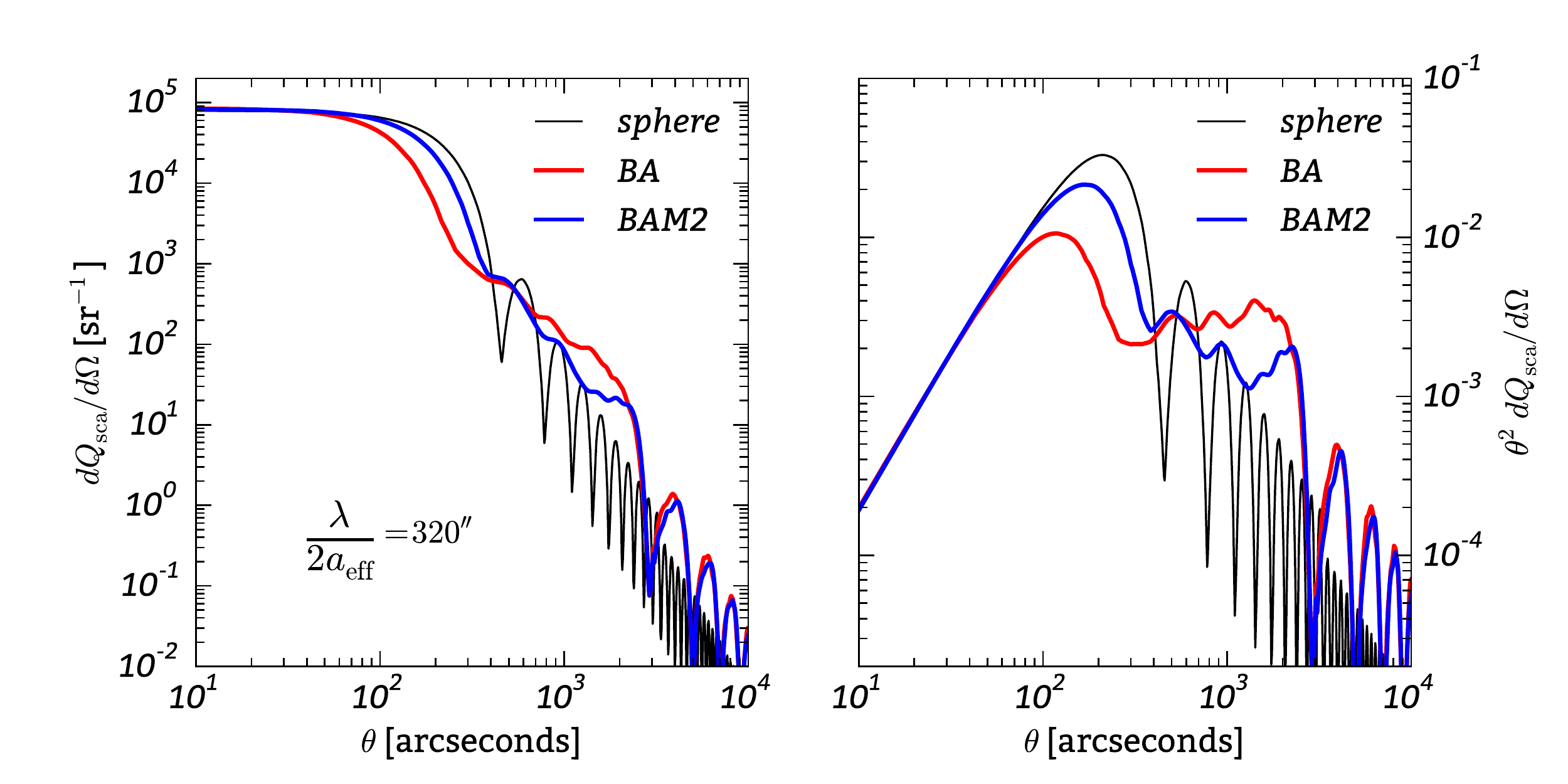}
\caption{\footnotesize\label{2dhm_1d} Azimuthally-averaged $dQ_{\rm
    sca}/d\Omega$ and $\theta^2dQ_{\rm sca}/d\Omega$ for the BA and
  BAM2 grains used in Figure \ref{2dheatmaps} (averaged over 300
  random orientations).
  Most of the scattered power is at $\theta\lesssim\lambda/2a_{\rm eff}=320\arcsec$.}  
\end{figure}

GGADT also calculates the differential scattering cross sections
$dC_{\rm sca}/d\Omega\equiv \pi a_{\rm eff}^2 dQ_{\rm sca}/d\Omega$.
Figure \ref{2dheatmaps} shows full two-dimensional (2D)
results for $dQ_{\rm sca}/d\Omega$ for BA and BAM2
aggregates with $a_{\rm eff}=0.2\micron$, each
composed of $N=256$ astrosilicate monomers
(monomer radii $0.0315\mu{\rm m}$).
For each grain, the shadow function for one particular orientation is shown, 
together with the 2D scattering pattern for that orientation.
The BA and BAM2 aggregates have equal masses, but
the BAM2 structure is more compact, and
its central scattering peak, with $\Delta\theta \approx \lambda/D$
(where $D$ is a characteristic ``diameter''),
is noticeably broader than for the BA structure.
For the selected orientations, the BA and BAM2 aggregates are each
elongated along the \newedit{$\hat{\bf x}$} axis,
resulting in scattering patterns that are elongated in the
\newedit{$\hat{\bf y}$} direction.  

The right-hand column of Figure \ref{2dheatmaps} shows the 2D scattering
pattern averaged over 300 random (in 3 dimensions) orientations of
the grain.  The BAM2 grain continues to have a significantly broader central
peak than the BA grain. 

Figure \ref{2dhm_1d} shows the azimuthally-averaged
$dQ_{\rm sca}/d\Omega$ from the right column of Figure \ref{2dheatmaps},
together with $dQ/d\Omega$ for an equal-mass sphere.  As expected,
the sphere has most of the scattered power
at $\theta < \lambda/D=320\arcsec$, but the BAM2 and BA structures
have {\it narrower} central peaks, because they are more extended
(larger $D$).

The similarity of the BA and BAM2 results
at $\theta > 3000\arcsec$, with regularly-spaced maxima and minima,
is striking.
At these large scattering angles, $dQ_{\rm sca}/d\Omega$ is determined by
the high-spatial frequency portion of the shadow function $f(x,y)$ 
(see Eq.\ \ref{eq:snhat}), and this comes from the spherical monomers.
Since the BA and BAM2 structures considered here have the same sized
monomers, they have very similar $dQ_{\rm sca}/d\Omega$ at large
scattering angles.  In the ISM, we of course do not expect agglomerate
structures to be composed of single-sized monomers, and scattering
halos will not show such maxima and minima.

\section{Discussion
		\label{sec:discussion}}

Observations of X-ray absorption and scattering by the ISM provide a
valuable opportunity for testing dust models and measuring
interstellar abundances of heavy elements, both in the gas and solid
phases.  By investigating the near-edge fine structure of absorption
(extinction) edges, one can deduce information about the (1) geometry
(e.g., porosity) and (2) composition of interstellar grains. The
dependence of the X-ray scattering halo on both scattering angle and
photon energy also constrains the size, structure, and composition of
the dust population.

Efforts to infer the dust composition, size distribution, and grain
geometry (shape and porosity) rely on detailed comparison of
observations (e.g., energy-dependence of the extinction, especially
near absorption edges) with theoretical calculations for various grain
models.  In this paper we have tested the widely-used approach of
WAM2000 \citep{Wilms+Allen+McCray_2000} for calculating dust
attenuation.

The WAM2000 method, which ignores scattering, does allow the elemental
abundances in dust to be estimated with reasonable accuracy (errors
$\lesssim 10-15\%$). However, because scattering is neglected, the
WAM2000 model does \emph{not} provide accurate profiles of the
near-edge fine structure in the extinction, and therefore should not
be used when attempting to distinguish between different possible
chemical forms in which the element of interest may be present (e.g.,
using the Fe L edge to try to distinguish metallic Fe, various Fe
oxides, or Fe contained in silicates). Inaccurate modeling of the
energy dependence of extinction near absorption edges can easily lead
to incorrect conclusions regarding the compositions of interstellar
grains.

Fortunately, for X-ray energies, the energy-dependent extinction, as
well as scattering halos, can readily be calculated accurately using
anomalous diffraction theory (ADT).  For spheres, the computational
cost of accurate ADT calculations is trivial, effectively equal to
that of the WAM2000 approximations.  One important advantage of the
ADT approach is that it is readily extended to other grain shapes,
such as spheroids, clusters of spheres, or indeed whatever shape is of
interest to the researcher.

The authors have developed \newedit{and made available} an open source ADT
Fortran suite --- GGADT --- that can perform the necessary ADT
calculations for both integrated absorption, scattering, and
extinction cross sections, as well as for differential scattering
cross sections. GGADT utilizes several numerical techniques to make
the ADT calculations quite fast without loss of accuracy \newedit{(see
Appendices \ref{efficient_calc_ggadt} and \ref{efficient_calc_ggadt2})}.

Above we used GGADT to calculate absorption and scattering by spheres,
spheroids, and random aggregates of sphere, to show explicitly that
the near-edge absorption fine structure is sensitive to grain
geometry, in particular, porosity. The fact that grain geometry alone
can significantly alter absorption edge profiles adds another
complication to future studies of near-edge X-ray extinction fine
structure, but in principle also provides another way to obtain
information about the shapes and structures of interstellar grains.

Figures \ref{2dheatmaps} and \ref{2dhm_1d} show how
X-ray scattering halos are sensitive to the geometric structure of the grains.
As shown by \citet{Draine+Allaf-Akbari_2006}, aligned interstellar grains
are expected to produce noncircular X-ray scattering halos, with large
enough asymmetries to be measurable by {\it Chandra}.
Grains with significant porosity produce significantly narrower forward
scattering peaks than equal-mass nonporous grains.
\citet{Heng+Draine_2009} showed that the implied changes to the X-ray
scattering halo could be used to test whether interstellar grains are
primarily dense, compact structures versus porous ``fluffy'' aggregates.
This will be the subject of future investigations, to see what grain
geometries are compatible with observations of extinction and polarization
at optical wavelengths, and scattering at X-ray energies.

\section{Summary}

The principal conclusions of this paper are as follows:

\begin{enumerate}

\item Neither the WAM2000 approximation, nor the Rayleigh-Gans approximation, should be used to calculate near-edge fine structure, even for spherical grains.  Both of these methods introduce systematic errors, which can be large for grain sizes and X-ray energies of astrophysical interest (see Figure \ref{fig:qext_vs_a}).

\item At X-ray energies, anomalous diffraction theory (ADT) is highly accurate and can be used for spheres (where agreement with exact Mie theory is excellent) as well as for more realistic grain geometries.

\item For spheres, ADT extinction, absorption, and scattering cross sections are given by simple analytic formulae (\ref{eq:Q_{ext} for sphere}-\ref{eq:Q_{sca} for sphere}). For more general geometries, ADT calculations of absorption, scattering, and extinction cross sections require only evaluation of the integrals in Eq.\ (\ref{adt_sigmas_abs}-\ref{adt_sigmas_ext}), which is computationally straightforward.

\item ``Naive'' calculation of X-ray scattering halos for general geometries can be computationally demanding (both operations and memory) when high angular resolution is desired.  We implement a method that greatly increases the computational efficiency, allowing accurate and high-resolution calculation of scattering halos to be performed with modest requirements of memory and CPU time.

\item We make available GGADT, an efficient open-source code to calculate X-ray absorption and scattering by grains with general geometries.

\item We use GGADT to calculate X-ray scattering and absorption by spheres, spheroids, and random aggregates. The near-edge extinction vs.\ energy can differ significantly among grains with the same mass and composition, but different shape. Grain geometry must therefore be taken into consideration when seeking to deduce grain composition from observations of near-edge X-ray fine structure.

\item In the X-ray regime, for complex grain geometries (e.g., porous
  grains), estimates of absorption and scattering cross sections made
  using homogeneous spheres with an ``effective'' refractive index
  obtained from an ``effective medium theory'' (EMT) do not, in
  general, provide accurate results. The scattering tends to be
  underestimated, and the near-edge fine structure is not accurately
  reproduced (see Figure \ref{extvsgeom_abs}).

\item We use GGADT to calculate X-ray scattering halos for porous
  grains, extending previous work by \citet{Heng+Draine_2009}.  For
  fixed grain mass, increasing porosity leads to narrowing of the
  forward scattering peak and characteristic halo angle (see
  Figure \ref{2dhm_1d}).  Observations of X-ray scattering halos can,
  therefore, provide constraints on the structure of interstellar grains.

\end{enumerate}
\acknowledgements
We thank Brandon Hensley for stimulating discussion, and the anonymous referee for helpful comments.
This work was supported in part by NSF grants AST1008570 and AST1408723.

\begin{appendix}
\section{Brief Description of GGADT\label{ggadt_desc}}

GGADT is a Fortran 90 code, provided with a GNU autotools
\texttt{configure} script \footnote{Autotools refers to autoconf
  (\url{http://www.gnu.org/software/autoconf/}) and Automake
  (\url{http://www.gnu.org/software/automake/})}, which should work on
most Unix/Linux/BSD-based operating systems (this includes Mac \newedit{OSX}). The
full source code \newedit{and documentation (both pdf and html) are available at
\url{http://www.ggadt.org}}.

There are two fundamental calculations that GGADT can perform:
\begin{enumerate}
\item \emph{Total} cross sections of a given grain and grain composition over a range of energies. 
\item Differential scattering cross section of a given grain and grain composition. 
\end{enumerate}
Both calculations can be averaged over a number of orientations; users can control how the orientation averaging is performed; either by choosing random orientations, or by evenly dividing the orientations over euler angles, or by specifying a list of orientations over which to average the calculations.

Input to GGADT can be done either on the command line, in a parameter file, or both (though combining command line arguments and parameter files is not recommended).

\section{Example Calculations Using GGADT\label{examples_ggadt}}

To calculate the differential scattering cross section for a cluster of silicate spheres with effective radius $a=0.1\mu{\rm m}$, at an energy of 500 eV 

\begin{lstlisting}[breaklines]
ggadt --grain-geometry='spheres' --aeff=0.1 --ephot=0.5 --norientations=100 --agglom-file=[..]/BA.256.1.targ --material-file=[..]/index_silD03
\end{lstlisting}

Here, \texttt{[..]} should be replaced with filepaths to the locations 
of these files. Files that describe the geometry of BA/BAM1/BAM2 grains 
\newedit{are available online at \url{http://ggadt.org/additional_files.html}}
The \texttt{--material-file} parameter expects an 
``index'' file; several of these are also provided online at 
\newedit{the previous url.}

To calculate the total cross section of an ellipsoidal silicate grain with axis ratios $x\!:\!y\!:\!z = 1\!:\!2\!:\!3$ at energies close to the O K edge (520 - 560 eV), one can run

\begin{lstlisting}[breaklines]
ggadt --integrated --grain-geometry='ellipsoid' --aeff=0.1 --grain-axis-x=1 --grain-axis-y=2 --grain-axis-z=3 --ephot-min=0.52 --ephot-max=0.56 --material-file=[..]/index_silD03
\end{lstlisting}

There are of order 30 different parameters and flags that one can set in GGADT; descriptions of all of these are given in the documentation\footnote{\url{http://ggadt.org/invoking.html}}. An \texttt{example} directory contains two sample cases for running GGADT along with a python script for plotting the results.

\section{Efficient Calculation of Orientation-Averaged $dC_{\rm sca}/d\Omega$\label{efficient_calc_ggadt}}

The shadow function $f(x,y)$ is represented on a numerical grid. To
obtain $dC_{\rm sca}/d\Omega$, 
we require $S(\hat{n})$, \newedit{proportional to} the Fourier transform of $f$
(see Eq.\ \ref{eq:snhat}). 
Naively, one could calculate the orientation-averaged
$dC_{\rm sca}/d\Omega$ by obtaining the complex shadow function
$f(\vec{x})$ on a numerical grid of $N_x\times N_y$ points, then using
a 2-dimensional FFT to calculate $S(\theta, \phi)$, converting this to
$dC_{\rm sca}/d\Omega$, repeating for many orientations, and averaging
the results.

For orientation-averaging, however, we can sample a single azimuthal angle,
taking $\phi=0$ without loss of generality. In this case, Equation \ref{eq:snhat} simplifies to
\begin{eqnarray}
S(\theta,\phi=0)&=& \frac{k^2}{2\pi} \int_{-\infty}^{\infty} \int_{-\infty}^{\infty} e^{ikx\sin{\theta}}f(x, y)dxdy\\
     &=& \frac{k^2}{2\pi} \int_{-\infty}^{\infty} e^{ikx\sin{\theta}} \underbrace{\left[\int_{-\infty}^{\infty} f(x, y)dy\right]}_{g(x)}dx\\
     &=& \frac{k^2}{2\pi} \int_{-\infty}^{\infty} e^{ikx\sin{\theta}}g(x)dx\\
     &=& \frac{k^2}{2\pi} \bar{g}(\kappa),
\label{eq:snhat_simple}
\end{eqnarray}
where $\bar{g}$ is the Fourier transform of $g(x) \equiv \int_{-\infty}^{\infty} f(x, y)dy$, and $\kappa = -\frac{1}{2\pi}k\sin\theta$. 

\begin{figure}[h!]
\centering
\includegraphics[width=1.0\linewidth]{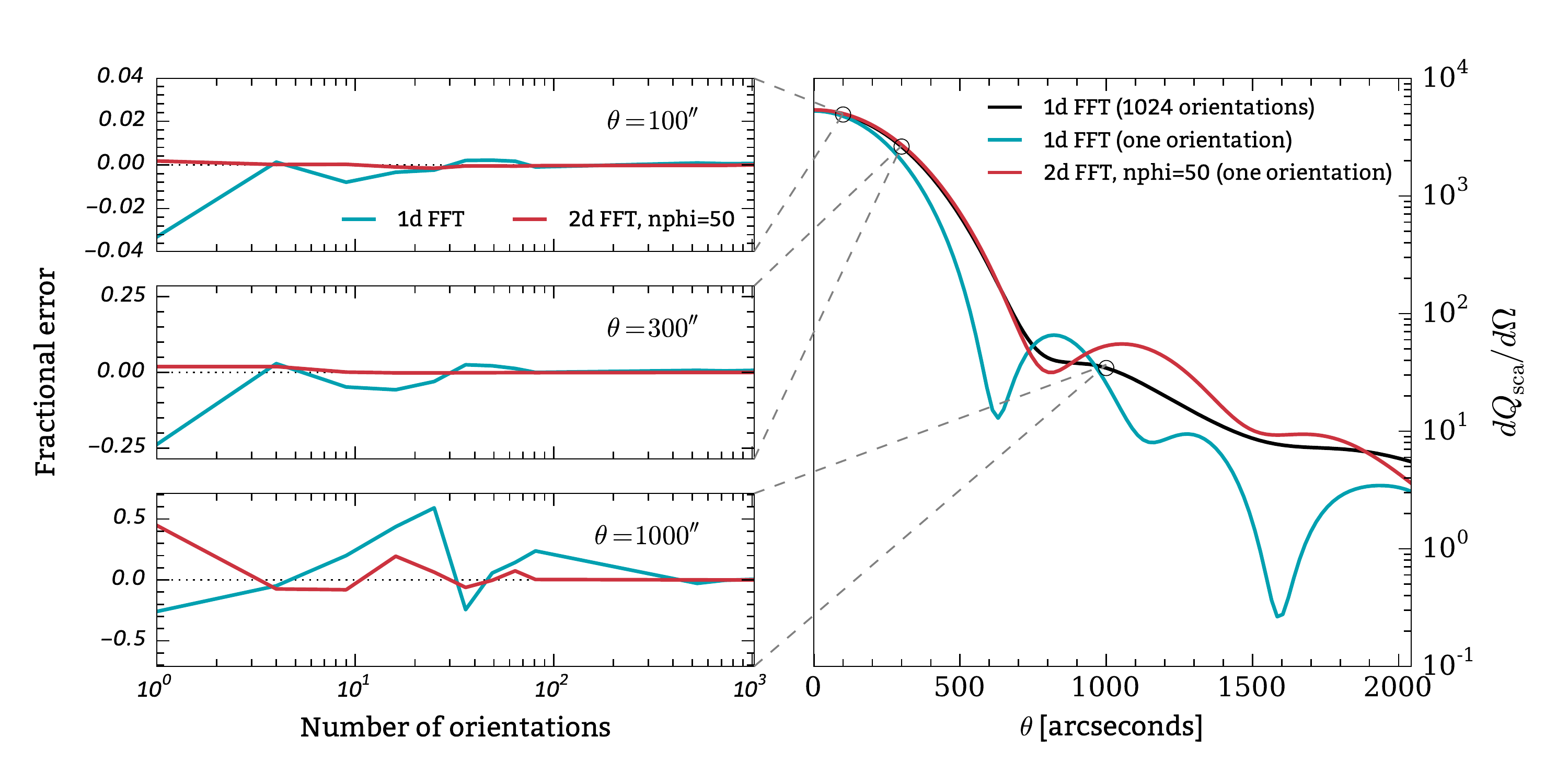}
\caption{\footnotesize\label{fig:norconv} Number of orientations
  needed for GGADT to converge to the true differential scattering
  cross section for $h\nu=2{\rm keV}$ and 
  $a_{\rm eff}=0.1 \mu{\rm m}$ silicate BAM2
  aggregate with 256 monomers (see Figure \ref{agglomgrains}).
  When using a
  2-D FFT and averaging over $\phi$, fewer orientations are
  needed to achieve the same accuracy as when 1-D FFT's are
  used. }
\end{figure}

By reducing a 2-dimensional calculation to a 1-dimensional
one, we speed up the calculation for a single orientation. The number
of operations required to obtain $S(\theta, \phi=0)$ is $\sim \O{N_y}
+ \O{N_x\log N_x}$, compared to $\sim\O{N_xN_y\log N_xN_y}$ for the
2-dimensional FFT. However, since we are approximating the
orientation-averaged differential scattering cross section by
averaging over a \emph{finite} number of orientations, the 1d FFT
trick will require more orientations to reach the same level of
accuracy as the 2d FFT, since we could sample many $\phi$ values with
a single 2d FFT. However, as shown in Figure \ref{fig:norconv}, only a
modest number of orientations ($\O{10^2}$) is needed to obtain
accurate numerical results. This means that the speedup gained from
using a 1d FFT can outweigh the cost of averaging over extra
orientations ($\sim\!10^2N_x \log N_x < \,\,\sim\!\!N_xN_y\log N_xN_y$).

An additional contribution to the computation 
time comes from the calculation of
the shadow function. For spheres and ellipsoids, the shadow function
computation time scales as $\O{N_xN_y}$. For agglomerations of
spheres, this process becomes approximately $\O{N_m^{1/3}N_xN_y}$,
where $N_m$ is the number of monomers.

The reason that aggregates require a $\O{N_m^{1/3}N_xN_y}$ computation
time is the following: the contribution to the shadow function from
each monomer is computed over a sub-grid of the full shadow function
grid, with $N_x'N_y'\approx (a_m/a)^2 N_xN_y$, where $a_m$ is the
radius of the monomer and $a$ is the effective radius of the
grain. Since $N_m a_m^3 = a^3$, we have that $a_m = a N_m^{-1/3}$ and
thus $N_x'N_y' = N_m^{-2/3}N_xN_y$. The computation time scales as
$N_m N_x'N_y' = N_m^{1 - 2/3}N_xN_y = N_m^{1/3}N_xN_y$.

For custom grain geometries, for which no assumptions about the structure
of the grain can be made, computation of the shadow function will
scale as $N_xN_yN_z$, where $N_z$ is the number of samplings along the
$\hat{z}$ direction. In this case, computation of the shadow function
dominates the computation time, and thus 2d FFT's are the more
efficient solution. In the former cases, (for clusters of spheres, as
long as $N_m \ll N_xN_yN_z$), the FFT dominates the computation time,
and thus 1d FFT's are more time efficient.

\section{Efficient calculation of $dC_{\rm sca}/d\Omega$ for high angular resolution\label{efficient_calc_ggadt2}}
Recall that the (one-dimensional) discrete Fourier transform (DFT) of an array of values $f_n$ is given by
\begin{eqnarray}
\hat{f}_m &=& \sum_{n=0}^N e^{-2\pi i m \frac{n}{N}} f_n\\
      &=& \sum_{n=0}^N  e^{-2\pi i m \frac{x_n - x_0}{L}}f_n\\
      &=& e^{2\pi imx_0/L}\sum_{n=0}^N e^{-2\pi i \left(\frac{m}{L}\right)x_n}f_n
\end{eqnarray}

Assume (for the sake of simplicity) that $N_x = N_y = N_z = N$. Here,
$\Delta x$ is the width of a grid element and $N$ is the total number of
grid elements along one direction.
When we calculate the DFT of $g(x)$, we obtain
$S(\theta_m, \phi=0)$ for a set of angles $\theta_m$, where
\begin{equation}
\sin\theta_m = \frac{2\pi m}{Nk\Delta x}.
\end{equation}

In astronomical contexts, usually only small angles ($\theta_{\rm
  max}\lesssim 10^4$ arcseconds) are relevant, and thus $\sin\theta_m
\approx \theta_m$. The smallest (non-zero) scattering angle for which
we can compute $dC_{\rm sca}/d\Omega$ is
\begin{equation}
\theta \approx \frac{2\pi}{kL} = \frac{\lambda}{L},
\end{equation}
where $L=N\Delta x$ is the extent of the grid. If $L=$ diameter
of the target, then this angle is approximately the size of the
central scattering lobe. Generally, we wish to resolve the central
lobe with resolution
\begin{equation}
\delta\theta_{\rm sca} = \frac{2\pi/kL}{N_{\rm sca}},
\end{equation}
where $N_{\rm sca}$ is the number of angles sampling the central scattering peak.

For a fixed grid resolution $\Delta x$ and photon wavenumber $k$, we
can increase the resolution of the differential scattering cross
section at small angles by extending the $xy$ grid on which the shadow
function $f(x,y)$ is defined, and setting $f(x, y)=0$ beyond the
actual shadow. If $L_{\rm ext}$ is the size of this extended grid,
then the DFT will have resolution
\begin{equation}
\Delta \theta \approx \frac{2\pi}{kL_{\rm ext}}.
\end{equation}
This technique is also known as ``padding.'' However, padding is a memory-intensive (and time-intensive) process. We seek a more efficient algorithm that provides high resolution at small scattering angles and avoids calculating $S(\theta)$ for $\theta > \theta_{\rm max}$.

One can throw away calculations of $\theta > \theta_{\rm max}$ by performing what is known as a ``pruned FFT.''\footnote{See \url{http://www.fftw.org/pruned.html} for more information.} Using a pruned FFT changes the asymptotic execution time of a one-dimensional FFT from $\O{N\log N}$ to $\O{N\log K}$, where $K$ is the largest index of the DFT that we care about; in this context, $K$ corresponds to the $m$ value of $\theta_{\rm max}$:
\begin{equation}
K \approx \frac{Nk\Delta x\sin\theta_{\rm max}}{2\pi}.
\end{equation}

Utilizing a pruned FFT \newedit{does not} speed things up very much on its own unless $K \ll N$, and we are still stuck with needing to pad the grid to increase the angular resolution of $dC_{\rm sca}/d\Omega$. There is, however, another way of improving the angular resolution of $dC_{\rm sca}/d\Omega$ that does not involve extending (padding) the shadow function grid. 

Instead of performing a single DFT on an extended (padded) grid, we can perform several \emph{offset} DFT's on the unpadded grid. A single DFT on the unpadded grid gives a coarse sampling of $dC_{\rm sca}/d\Omega$ over angles $\theta_m$. We can then perform another DFT to produce another coarse sampling of $dC_{\rm sca}/d\Omega$ over \emph{a different set of angles}, $\theta_{m + \delta}$. We've now doubled the angular resolution of $dC_{\rm sca}/d\Omega$ by performing 2 unpadded DFT's. This process takes advantage of the following:
\begin{eqnarray}
\hat{f}_{m + \delta } &=& \sum_{n=0}^N e^{-2\pi i (m+\delta )\frac{n}{N}} f_n\\
    &=& \sum_{n=0}^N e^{-2\pi i m\frac{n}{N}} \underbrace{ \left(e^{-2\pi i \delta \frac{n}{N}} f_n\right) }_{F_n}\\
    &=& \widehat{F}_m.
\end{eqnarray}

By doing offset DFT's of $g$ for as many $\delta$ values as necessary
to achieve the desired resolution in $\theta$, we avoid the need for
any padding. The complexity of this entire algorithm is $\O{N_{\rm
    FFT}N\log{K} + \beta N^2}$, where the $\beta N^2$ term comes from
the calculation of $g(x)$ (scaled by some constant $\beta = \O{1}$),
and $N_{\rm FFT}= (\theta_{\rm max}/\delta\theta_{\rm sca}) / K$ is
the number of FFT's necessary to achieve the desired resolution in the
scattering angle, $\delta \theta$.

The naive calculation of $S(\theta)$, using the full two-dimensional
FFT with a padded grid, has a computational time $t_{\rm naive}
\propto N_{\rm pad}^2\log N_{\rm pad}$, where $N_{\rm pad} =
2\pi/(k\Delta x \delta \theta_{\rm sca} )$. Compare this to the unpadded grid,
where $N \approx 2a/\Delta x$, where $a$ is the characteristic radius
of the grain. We have that
\begin{equation}
t_{\rm naive}\propto \left(\frac{N\pi}{ka\delta\theta_{\rm
    sca}}\right)^2\log\left(\frac{N\pi}{ka\delta\theta_{\rm
    sca}}\right).
\end{equation}

\begin{figure}[h!]
\centering
\includegraphics[width=15.0cm,clip=true,
                 trim=0.5cm 0.2cm 0.5cm 1.0cm]
{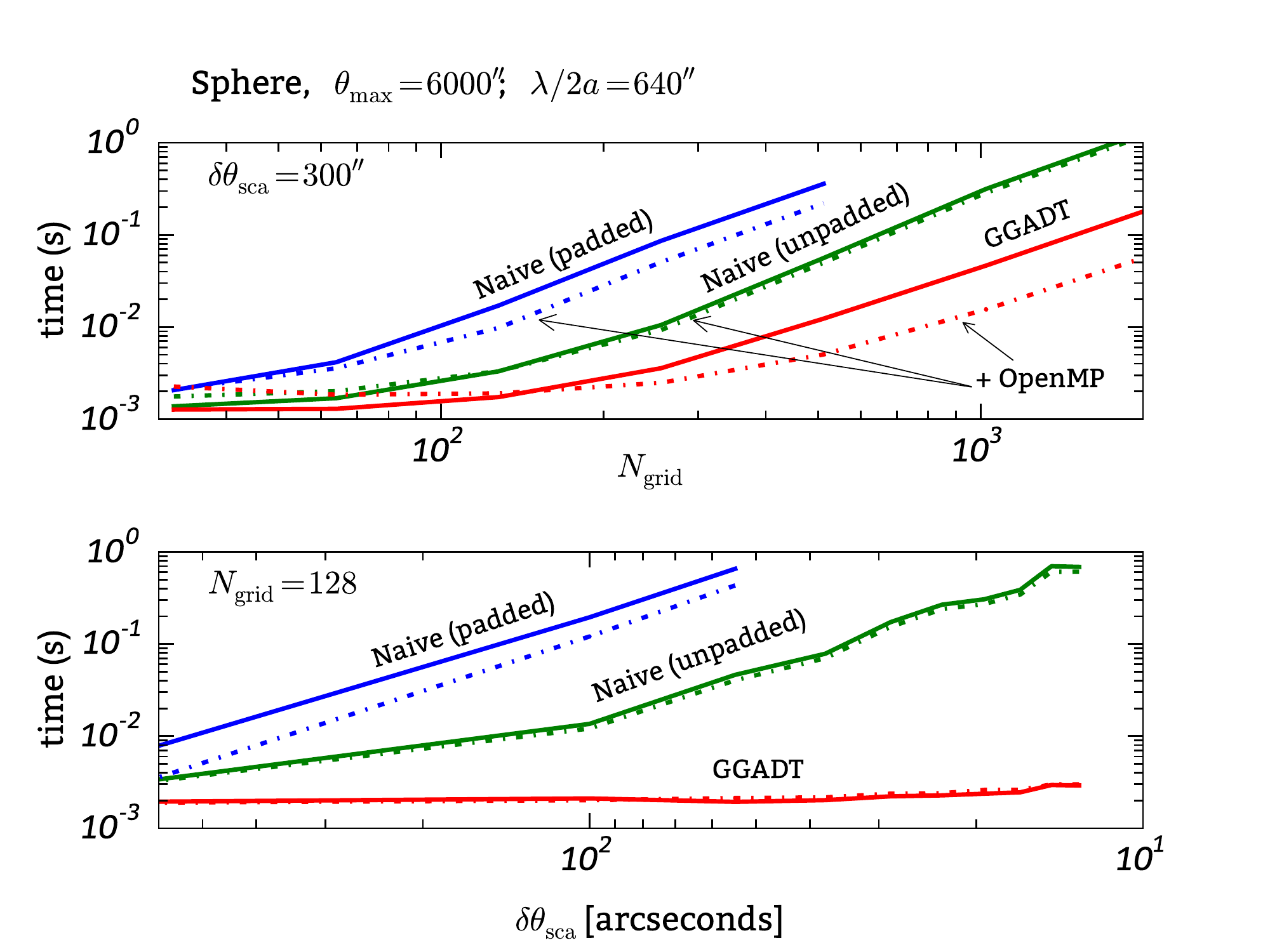}
\caption{\footnotesize\label{fig:timing} GGADT execution times as a
  function of grid size $N_{\rm grid}$ (upper panel) and desired
  angular resolution $\delta\theta_{\rm sca}$ (lower panel).  Times
  are for system with one 2.6GHz Intel Core i7 processor with 4 cores.
  Also shown are times for more naive methods of calculating $dC_{\rm
    sca}/d\Omega$ (see text).  Solid lines are for serialized version
  of GGADT (no parallel computation). Dash-dotted lines are for OpenMP, 
  using 4 threads. GGADT timing is almost independent of
  $\delta \theta_{\rm sca}$ and improves upon the padded
  two-dimensional FFT calculation by a factor $\sim$30 for $N_{\rm
    grid}\times N_{\rm grid}$ grids with $N_{\rm grid}\approx 500$.
  Test problem has $\lambda/\Delta x=320\arcsec$ (e.g., $\Delta
  x=0.4\micron$ and $E=2$keV) where $\Delta x$ is the maximum linear
  extent of the shadow function and $\theta_{\rm max}=6000\arcsec$.}
\end{figure}

Utilizing multiple FFT's [$N_{\rm FFT} = (\theta_{\rm
    max}/\delta\theta_{\rm sca}) / K$] of length $K = N k\Delta x(\theta_{\rm
  max}/2\pi) = x(\theta_{\rm max}/\pi)$, requires a computational time
that scales as
\begin{equation}
t_{\rm GGADT} \propto \left(\frac{\pi}{ka\delta\theta_{\rm sca}}\right)\log\left(ka\frac{\theta_{\rm max}}{2\pi}\right) + \beta N^2.
\end{equation}
We have turned an $N^2\log N$ problem into an $N^2$ problem, and, as
shown in Figure \ref{fig:timing}, GGADT is approximately two orders of 
magnitude faster than calculations that use a padded two-dimensional FFT.

Another key advantage of GGADT over naive padding methods lies in
memory requirements. Padded FFT's require storing a 2-dimensional
array of floats, with dimension $N_{pad}\times N_{pad}$. This
corresponds to a memory requirement of
\begin{equation}
M_{\rm naive} = 1.37~{\rm GB}\left(\frac{M_0}{8~{\rm
    bytes}}\right)\left(\frac{100''}{\delta\theta_{\rm sca}}\right)^2\left(\frac{500~{\rm
    eV}}{E}\right)^2\left(\frac{0.1\mu{\rm m}}{a}\right)^2
\left(\frac{N}{512}\right)^2,
\end{equation}
\refereenote{Changed ``Gb'' (gigabit) to GB (gigabyte) and 16 bytes to 8 bytes for a \texttt{double} (and 
divided the 2.74 GB number by two accordingly).}
where $M_{\rm naive}$ is the total memory required to store the array,
and $M_0$ is the memory requirement for a single floating point number
(we assume \newedit{8} bytes, the standard length of a double-precision
floating point number). For a fiducial case of $500$ eV, $a =
0.1\mu{\rm m}$, $\delta\theta = 100''$, and $N = 512$, we have that
\newedit{$M_{\rm naive} \approx 1.37$} GB, a hefty memory
requirement even for computers at the time of this writing. GGADT, on
the other hand, requires a mere \newedit{$M_0 N^2 = 2\times 10^6 (N/512)^2$}
bytes. It's possible to improve this to \newedit{$M = M_0 N \approx 4 \times
10^3(N/512)$} bytes by avoiding storing the entire 2-dimensional shadow
function.

\end{appendix}
\bibliography{btdrefs,jahrefs}

\end{document}